\documentclass[10pt,journal,compsoc]{IEEEtran}

\ifCLASSOPTIONcompsoc
  \usepackage[nocompress]{cite}
\else
  \usepackage{cite}
\fi

\ifCLASSINFOpdf
  \usepackage[pdftex]{graphicx}
\else
  \usepackage[dvips]{graphicx}
\fi

\usepackage{booktabs} 
\usepackage{float}
\usepackage{pifont}
\usepackage{color}
\usepackage[ruled,vlined]{algorithm2e}
\usepackage{algorithmic}
\usepackage{multirow}
\usepackage{amsmath,amssymb}
\usepackage{bbm}
\usepackage{mathrsfs}
\usepackage{array,multirow}
\usepackage{caption}
\usepackage{subcaption}
\usepackage[hang]{footmisc}

\newcolumntype{C}[1]{>{\centering\arraybackslash}m{#1}}

\graphicspath{{./Figures/}}

\DeclareMathOperator*{\argmin}{argmin}

\begin{document}

\title{Man-in-the-Middle Attacks against Machine Learning Classifiers via Malicious Generative Models}

\author{Derui~(Derek)~Wang,
        Chaoran~Li,
        Sheng~Wen,
        Surya~Nepal,
        and~Yang~Xiang
\IEEEcompsocitemizethanks{\IEEEcompsocthanksitem {D. Wang, C. Li, S. Wen, and Y. Xiang} are with School of Software and Electrical Engineering, Swinburne University of Technology, Hawthorn, VIC 3122, Australia;\protect\\
E-mail: \{deruiwang, chaoranli, swen, yxiang\}@swin.edu.au;
\IEEEcompsocthanksitem S. Nepal is with CSIRO's Data 61, Australia; \protect\\
E-mail:surya.nepal@data61.csiro.au;
\IEEEcompsocthanksitem D. Wang and C. Li are also with CSIRO's Data 61, Australia;
\IEEEcompsocthanksitem D. Wang and C. Li contributed equally to this paper.}

\thanks{Manuscript received XX XX, 2019; revised August 30, 2019.}}

\markboth{Submitted to IEEE TDSC, October 2019}%
{Shell \MakeLowercase{\textit{et al.}}: Bare Demo of IEEEtran.cls for Computer Society Journals}

\IEEEtitleabstractindextext{%
\begin{abstract}
Deep Neural Networks (DNNs) are vulnerable to deliberately crafted adversarial examples. In the past few years, many efforts have been spent on exploring query-optimisation attacks to find adversarial examples of either black-box or white-box DNN models, as well as the defending countermeasures against those attacks. In this work, we explore vulnerabilities of DNN models under the umbrella of Man-in-the-Middle (MitM) attacks, which has not been investigated before. From the perspective of an MitM adversary, the aforementioned adversarial example attacks are not viable anymore. First, such attacks must acquire the outputs from the models by multiple times before actually launching attacks, which is difficult for the MitM adversary in practice. Second, such attacks are one-off and cannot be directly generalised onto new data examples, which decreases the rate of return for the attacker. In contrast, using generative models to craft adversarial examples on the fly can mitigate the drawbacks. However, the adversarial capability of the generative models, such as Variational Auto-Encoder (VAE), has not been extensively studied. Therefore, given a classifier, we investigate using a VAE decoder to either transform benign inputs to their adversarial counterparts or decode outputs from benign VAE encoders to be adversarial examples. The proposed method can endue more capability to MitM attackers. Based on our evaluation, the proposed attack can achieve above 95$\%$ success rate on both MNIST and CIFAR10 datasets, which is better or comparable with state-of-the-art query-optimisation attacks. At the meantime, the attack is $10^4$ times faster than the query-optimisation attacks.
\end{abstract}

}

\maketitle

\IEEEdisplaynontitleabstractindextext

%
\IEEEpeerreviewmaketitle

\IEEEraisesectionheading{\section{Introduction}\label{sec:introduction}}

%
%
%
%
\IEEEPARstart{D}{}eep Neural Networks (DNNs) are powerful tools for Machine Learning (ML) tasks. However, compared to other ML models, DNNs are more vulnerable towards adversarial attacks such as adversarial examples \cite{Goodfellow2014Explaining} and back-door attacks \cite{ji2018model}. As one major threat to ML security, the problem of adversarial example was first investigated in \cite{Szegedy2013Intriguing}. Adversarial examples denote data examples that are quasi-imperceptibly perturbed to achieve particular adversarial objectives (\textit{e.g.}, misclassification of ML classifiers and mis-detection of object detectors). Existing methods for crafting adversarial examples are mostly based on a query-optimised framework. Given a data example, an attacker needs to query a target DNN model to calculate an adversarial cost/fitness score. Based on the adversarial cost/fitness score, the attacker can then adopt optimisation techniques (\textit{e.g.}, gradient descent, or zeroth-order optimisation methods) to adjust the feature values of the data example to make it an adversarial one.

On the other hand, Man-in-the-Middle (MitM) attack is one of the most lethal cyber threats \cite{desmedt2011man}. An attacker stealthily alters the communication between two parties and then deliver malicious payloads to the parties to achieve adversarial goals. In the field of machine learning, the vulnerability of DNN models has not been extensively studied within the scope of MitM attacks. A typical ML task that can be exploited by MitM attackers is classification/example labelling. For instance, in the case of smart manufacturing, camera-captured images are sent to either a local or an online DNN classifier for classification. Based on the classification results, possible security issues or faulty products can be detected and then resolved. ML applications taking the same spirit can also be discovered in the systems such as autonomous vehicles, and ML-as-a-service supported mobile devices. In these applications, an adversary can stand in between the collected data and the DNN classifier to launch MitM attacks. The attacks aim to fool the DNN model by stealthy manipulating the submitted data, such that the model generates incorrect outputs which may satisfy the malicious goal of the attacker.

For an MitM adversary, the aforementioned methods for crafting adversarial examples are not practically sound. First, these attacks are one-off, which means that the adversarial perturbations or the learnt experience cannot be directly applied on new examples to make them adversarial (\textit{i.e.}, the attacks are not data-agnostic). Second, instead of initialising cold-start attacks, these methods require the attacker to query the victim model multiple times beforehand. In practice, it might be impossible to query the victim model without alerting defences \cite{roschke2009intrusion}. There are attempts to craft universal perturbations to address the problems \cite{moosavi2017universal}. Otherwise, generative models such as Generative Adversarial Networks (GANs) have been trained to produce adversarial counterparts of benign inputs. Variational Auto-Encoder (VAE) is another important category of generative models. So far, VAEs are less noticed for generating adversarial examples. To our best knowledge, there are a few attacks that manipulate the latent variables of VAEs to lead to mis-generation of images\cite{kos2018adversarial}. However, VAE has not been used for generating adversarial examples of DNN classifiers.

In spite of the success of GANs, VAEs have several advantages compared to GANs. First, VAEs are much easier to be trained than GANs \cite{tolstikhin2018wasserstein}. Second, the latent variables encoded/decoded by a VAE are set to follow normal distributions \cite{brock2017neural}. Consequently, it is easy to generate new examples by using latent vectors sampled from the normal distributions as the inputs of a VAEs. Additionally, VAEs can easily control the characteristics of the generated examples by latent vector arithmetic. Based on the advantages, a malicious VAE can either transform a benign example to adversarial or generate novel adversarial examples from sampled latent variables. This flexibility can expand the attack vectors and the capability of an MitM attacker. An investigation of generating adversarial examples using VAEs might enhance the understanding towards adversarial examples. It will also foster the process of amending vulnerabilities of ML applications such that the applications become more trustworthy and secure. 

To study the above research questions, in this paper, we design a Malicious Variational Decoder (MVD) which can be concatenated to arbitrary VAE encoders to generate adversarial examples of benign inputs of the VAE encoders. The MVD is flexible as an MitM adversary. The MVD can either be used together with a VAE encoder to generate adversarial examples in real-time against DNN classifiers, or it can replace the decoders in DNN-based systems with a VAE-classifier structure (e.g. classifiers collecting data by wireless sensor networks \cite{liu2019cbn}) to maliciously decode the output from benign VAE encoders. To the best of our knowledge, we make the first attempt towards crafting adversarial examples using VAEs. We summarise our contributions as follows:

\begin{itemize}
\item\textit{We analyse the security properties of DNN classifiers from a perspective of MitM attackers;}
\item\textit{We propose launching cold-start MitM attacks in real-time by using VAEs to generate adversarial examples;}
\item\textit{We propose a method for training malicious variational decoder to generate adversarial examples on-the-fly, without accessing the outputs from the victim classifier beforehand;}
\item\textit{The malicious decoder can decode outputs from benign VAEs into adversarial examples;}
\item\textit{We discuss the possibility of extending the MitM attacks against the real-world applications of machine learning models.}
\end{itemize}

The paper is organised as follows: Section \ref{S::Background} provides background knowledge about adversarial examples and VAE. Section \ref{S::Related_work} introduces related work in the areas of adversarial example attacks. Section \ref{S::Primer} formulates the problem of MitM attacks and presents the threat models used in this paper. Section \ref{S::Method} introduces the detailed methods for building. In Section \ref{S::Eval}, we present the evaluation results. Section \ref{S::Discuss} discusses the proposed attack and possible defences towards the attack. Finally, Section \ref{S::Conclude} summarises the paper and concludes our future work.

\section{Background}\label{S::Background}
\subsection{Adversarial examples for ML classifiers}
Given a DNN classifier $f$, an adversarial example $\hat{x}$ of its counterpart benign example $x$ aims at achieving the following objective:
\begin{align}\label{E1}
&f(\hat{x}) \neq f(x) \\
&s.t.\ \|\hat{x} - x\|_p \leq \epsilon.\nonumber
\end{align}

Herein, $\|\cdot\|_p = (\sum_{i=1}^n|\cdot_i|^p)^{\frac{1}{p}}$ is a $p$-norm measurement on the distortion of the adversarial perturbation $\hat{x} - x$, and $\epsilon$ is a small value acting as an upper bound of the distortion. $\epsilon$ is also called the \texttt{budget} of the adversarial perturbation. The budget restricts the distortion on $\hat{x}$ to make it quasi-imperceptible for human. Directly optimising the objective in Equation \ref{E1} is difficult. Alternatively, an attacker define an adversarial loss function/fitness score $J$ to measure the progress towards having $f(\hat{x}) \neq f(x)$:
\begin{align}\label{E2}
& J(f(\hat{x}), f(x)) \\
&s.t.\ \|\hat{x} - x\|_p \leq \epsilon.\nonumber
\end{align}

Gradient descent techniques can be used to optimise the above objective in white-box settings. Otherwise, zeroth-order optimisation techniques (\textit{e.g.}, particle swarm, genetic algorithm, differential evolution, etc.) are employed for solving the optimisation in black-box settings.

\subsection{Variational autoencoder}\label{S:vae}
A VAE contains an encoder and a decoder parameterised by $\phi$ and $\theta$, respectively. Given a set $X$ of data examples, the VAE maximises the log-likelihood $\log p(x), x\in X$ based on observations. The encoder portion of the VAE models a posterior $p_{\phi}(z|x)$ of latent variables $z$ conditioning on $x$. Suppose the real conditional probability of the latent variable $z$ obeys a distribution $p_d(z|x)$. The objective of the VAE encoder is thus to minimise the distance between $p_{\phi}(z|x)$ and $p_d(z|x)$. The distance is measured by the Kullback-Leibler (KL) divergence between $p_{\phi}(z|x)$ and $p_d(z|x)$, that is:
\begin{equation}\label{E3}
KL[p_{\phi}(z|x)||p_d{(z|x)}] = -\int{p_{\phi}(z|x)\log{\frac{p_d{(z|x)}}{p_{\phi}(z|x)}}}dz \geq 0.
\end{equation}
Based on Bayes theorem, we can induce Inequality (\ref{E3}) to:
\begin{equation}\label{E4}
-\int{p_{\phi}(z|x)[\log{\frac{q_{\theta}{(x|z)p(z)}}{p_{\phi}(z|x)}}-\log p(x)}]dz \geq 0,
\end{equation}
wherein, $q_{\theta}{(x|z)}$ is actually modelled by the VAE decoder parameterised by $\theta$. We can further induce the above inequality to:
\begin{equation}\label{E5}
\log p(x) \geq -KL[p_{\phi}(z|x)||p(z)] + E_{\sim p_{\phi}(z|x)}{\log q_{\theta}(x|z)}.
\end{equation}

The right side of the above inequality is called the Evidence Lower Bound (ELBO) of $\log p(x)$. Since $p(x)=\int{p(x|z)p(z)dz}$ is intractable, the VAE maximises the ELBO instead of directly maximising the log-likelihood $\log p(x)$. The first term of the ELBO measures the distance between the conditional probability of the latent variables $z$ modelled by the encoder and the real latent distribution $p(z)$, and the second term is actually the reconstruction quality of the VAE. Herein, since $p(z)$ is an arbitrary distribution, we can define it as a standard Gaussian distribution $\mathcal{N}(0,1)$ for simplicity. Similarly, we can set $p_{\phi}(z|x)$ as a Gaussian distribution $\mathcal{N}(\mu,\delta)$. Therefore, we can compute the KL divergence between $p_{\phi}(z|x)$ and $\mathcal{N}(0,1)$ as the first term of the ELBO. We can re-parameterise the ELBO by $\mu$ and $\delta$ as follows:
\begin{equation}\label{E6}
ELBO = \frac{1}{2}{[1+\log(\delta^2)-\delta^2-\mu^2]} + E_{\sim p_{\phi}(z|x)}{\log q_{\theta}(x|z)}.
\end{equation}
Hence, training the VAE is actually minimising the negative of Equation (\ref{E6}). Given the dataset $X$, the objective cost function to minimise can then be written as:
\begin{align}\label{E7}
L = &-\frac{1}{\|X\|}\sum_{x\in X}\sum_{j=1}^{\|z_x\|}{\frac{1}{2}[1+\log(\delta_j^2)-\delta_j^2-\mu_j^2]}\\
&\ -\frac{1}{\|X\|}\sum_{x\in X}E_{\sim p_{\phi}(z_x|x)}{\log q_{\theta}(x|z_x)}\}.\nonumber
\end{align}
Notice that minimising the second term in the cost function is actually minimising the reconstruction error of the VAE. In practice, the VAE encoder determines the mean $\mu$ and the variance $\delta^2$ of $\mathcal{N}(\mu,\delta)$, and the decoder takes latent variables $z$ sampled from $\mathcal{N}(\mu,\delta)$ and then generate data examples.

\section{Related work}\label{S::Related_work}
Previous research about adversarial examples focuses on attacks based on first-order information of DNNs. In white-box settings, adversarial examples can be generated by either gradient-based methods or forward-derivative-based methods. The gradient-based methods compute adversarial gradients based on adversarial objective functions and update a data example according to the adversarial gradients, such that the example becomes its adversarial counterpart \cite{Goodfellow2014Explaining, Szegedy2013Intriguing, kurakin2016adversarial, Moosavidezfooli2016DeepFool, carlini2017}. Forward-derivative-based attack perturbs salient features based on the Jacobian between the model inputs and outputs \cite{Papernot2016The}. Additionally, there are evolved attacks that use different distortion metrics to make adversarial examples that are more imperceptible to human eyes \cite{hosseini2018semantic, engstrom2017rotation}. Furthermore, there are methods crafting adversarial examples of data that have discrete features (e.g. text) \cite{Jia2017, Ebrahimi2018, Grosse2017a}.

However, accessing the first-order derivatives of the target DNNs is less practical in real-world attacks. As an enhancement, Zeroth-Order (ZO) optimisation techniques can be used in attacks when the victim models are black-boxes to an attacker. \cite{nitin2018practical, ilyas2018black}, and \cite{liu2018zeroth} employed gradient estimate techniques to find adversarial examples of black-box classifiers. Particle swarm optimisation was used in \cite{sharif2016accessorize} to attack black-box face recognition models. Similarly, \cite{alzantot2018genattack} adopted genetic algorithm to attack DNN models. The ZO attacks mainly focus on reducing query complexity and minimising required information during the attacks. However, the attacker still needs to obtain the output from the DNN models. Moreover, the attack is not data-agnostic. 

On the other hand, using generative models to craft adversarial example can launch data-agnostic attacks without accessing the output of the DNN models. Adversarial examples have not been extensively researched together with generative models. In recent research, adversarial images that result in mis-generation of VAEs are studied in \cite{kos2018adversarial}. A GAN used to transform benign examples to adversarial examples is proposed in \cite{xiao2018generating}.

\section{Primer}\label{S::Primer}
\subsection{Problem definition}\label{S::Problem_def}
In this study, we first study using VAE to transform benign inputs into adversarial ones. Second, we consider MitM attacks against applications that have a VAE-classifier structure. That is, a VAE is inserted between the raw input data and a classifier. This structure can often be seen in applications such as data denoising \cite{vincent2008extracting,creswell2018denoising} anomaly detection \cite{an2015variational}, and collaborative feature extraction \cite{li2017collaborative}. To minimise the changes that need to be made by the attacker, we suppose that the attacker can only alter the parameters of the decoder to result in misclassification of the classifier. The encoder $e_{\phi}$ parameterised by $\phi$ is kept benign, and it conventionally encodes the inputs into the latent space.

Based on the clarification of the problem, we derive a formal definition of the problem as follows:

\textbf{Definition 1}: \textit{Given a victim model $f$, a benign input example $x$, the problem of man-in-the-middle attack can be interpreted as $f(g_{\theta}(x)) \neq f(x)$. Herein $g_{\theta}$ is an adversarial transformation in between the input and the model. Otherwise, the attacker cannot alter $x$ by any form of data injection. The attacker's objective is to find the best $\theta$ that results in the adversarial consequence (\textit{i.e.}, misclassification).}

\begin{figure}[t!]
\center
\includegraphics[width=1\linewidth]{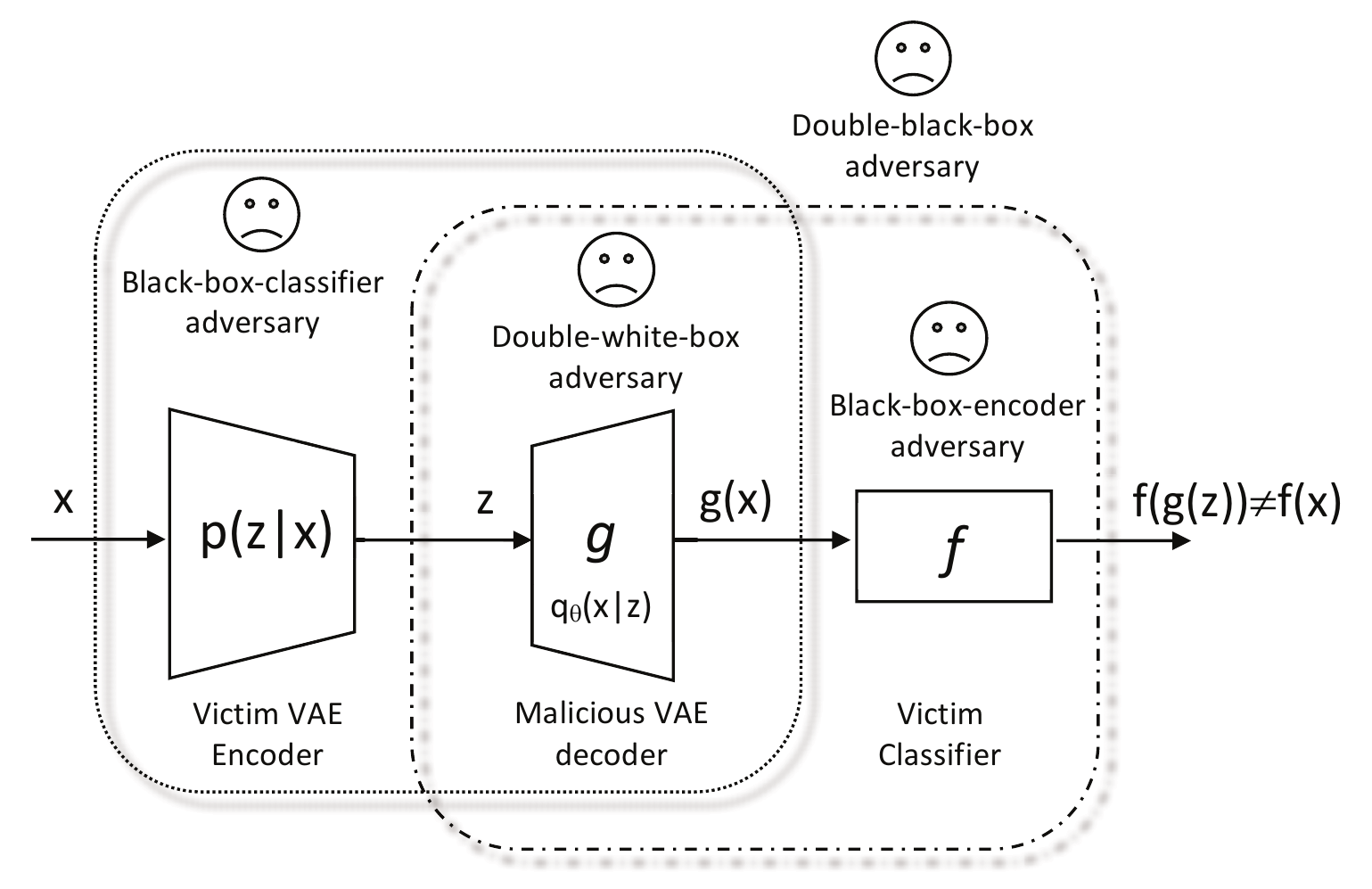}
\caption{Illustration of the problem definition and threat models of MitM attacks in this paper. An adversary attacks via manipulating the decoder in between a VAE encoder and a classifier.}
\label{p_def}
\end{figure}

To investigate whether the defined problem is achievable, we first need to bound the \textit{a priori} knowledge of $f$ and $x$ obtained by $g_{\theta}$. We thus model the problem in different settings in the following section.

\subsection{Threat models} \label{S::Threat_model}
In this section, we define the threat models of the adversaries that launch man-in-the-middle attacks towards machine learning models with the VAE-classifier structure. As a generalisation, we model the threat in both black-box and white-box settings. Based on the settings of the VAE encoder and the classifier, we propose four threat models in total.

In the first threat model, the attacker $g_{\theta}$ has every access to the victim classifier $f$, the VAE encoder $e_{\phi}$ and a training dataset $\ddot{X}$ that comes from a similar distribution with the one modelled by the classifier. This defines a \texttt{double-white-box} setting. In this setting, $g_{\theta}$ can launch attacks directly based on the latent distribution modelled by the encoder $e_{\phi}$, and it optimise the parameters $\theta$ according to the first-order information $f'(g_{\theta}(x))$ of the victim classifier $f$. 

In a \texttt{double-black-box} setting, $g_{\theta}$ can only query $f$ rather than accessing the gradients (\textit{i.e.}, the first-order information) of $f$. Furthermore, the attacker does not have the distribution of the latent variable beforehand, and the attacker is not allowed to estimate the latent distribution by querying the encoder. However, the attacker should have access to the dimensionality of the VAE encoder output and the dataset $\ddot{X}$. In this setting, first, the attacker is allowed to train a shadow VAE encoder to estimate the latent distribution. Second, a shadow model $\ddot{f}$ can be training locally by the attacker to estimate $f$. $g_{\theta}$ can use the gradients $\ddot{f}'{g_{\theta}}$ to achieve an adversarial objective $\ddot{f}(g_{\theta}(x)) \neq \ddot{f}(x)$

If $g_{\theta}$ in the double-black-box setting can access $e_{\phi}$ as a white-box, this defines a \texttt{black-box-classifier} setting. In the contrast, if $g_{\theta}$ can access the classifier as a white-box, this is a \texttt{black-box-encoder} setting. As a conclusion, we illustrate the problem definition along with the threat models in Figure \ref{p_def}.

\section{Design of the attack}\label{S::Method} 
In this section, we first resolve the problem of launching attacks in the double-black-box setting. Subsequently, we derive the adversarial loss functions and the optimisation techniques for training a malicious variational decoder. At last, we propose a general framework for training the malicious variational decoder.

\begin{figure*}[t!]
\center
\includegraphics[width=1\linewidth]{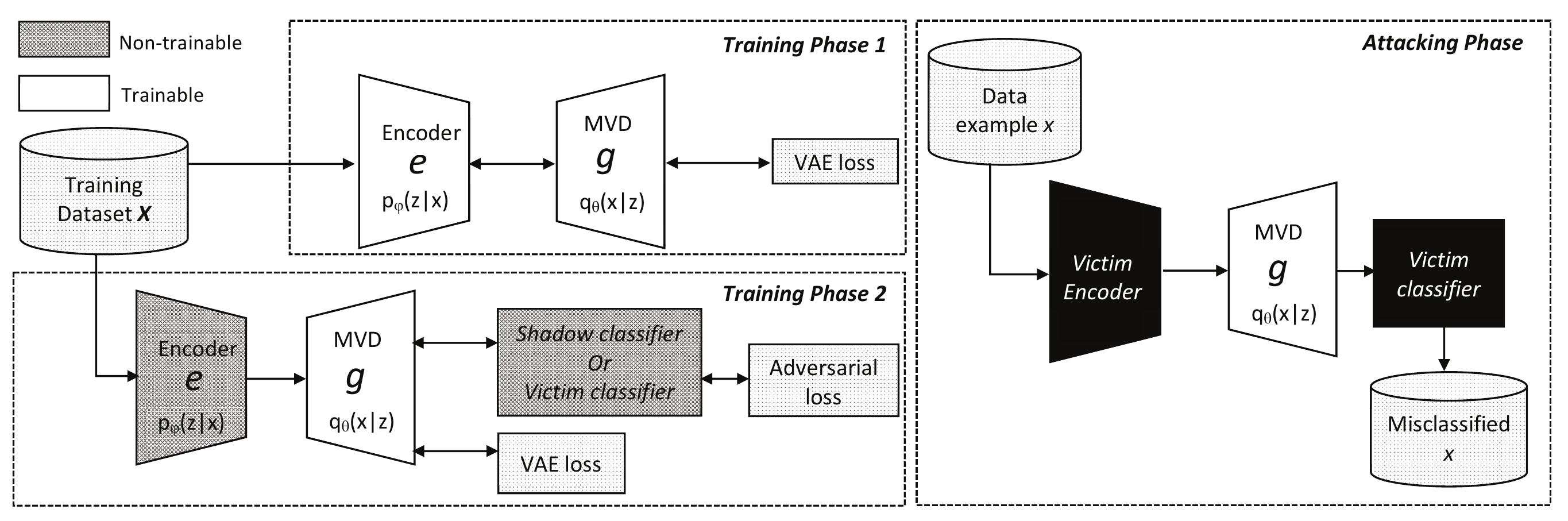}
\caption{The framework of training the MVD and launching attacks. We build the MVD via a two-step training process. We first train a benign VAE following the standard subroutine in step one. In the second step, we first calibrate the decoder such that it can decode from a black-box encoder. We then train the decoder to an MVD based on an adversarial loss function. In the attacking phase, the attacker can insert the decoder in between an arbitrary VAE encoder and a victim classifier to result in misclassification of the classifier.}
\label{framework}
\end{figure*}

\subsection{Calibrating black-box attacks}\label{S::Method::1}
The definition of launching double-black-box MitM attacks against VAE encoders and DNN classifiers is twofold: 1) the attacker can attack black-box classifiers, and 2) the attacker can launch attacks without knowing the parameters of the victim VAE encoders. Herein, the second case actually models an MitM attack against VAE-classifier structured models. When the victim classifier is a black-box to the attacker, a commonly adopted practice is to train a shadow model that estimates the victim classifier. Later on, the shadow model can be connected to the MVD to provide estimated adversarial gradients of the victim model. Therefore, the remaining question is, whether the MVD can attack black-box VAE encoders?

Given any arbitrary VAE encoder that models a distribution $p_{*}(z|x)$, the MVD attempts to reconstruct a adversarial example $\hat{x}$ that achieves the same consequence (\textit{i.e.}, resulting in either non-targeted or targeted misclassification from $f$) without exceeding the perturbation budget $\epsilon$. Notice that given a training dataset $X$, the optimised ELBO can be re-written as follows:
\begin{equation}\label{eq1}
-E_{x\sim X}KL[p_{\phi}(z|x)||p(z)] + E_{x\sim X}E_{\sim p(z)}{\log q_{\theta}(x|z)}
\end{equation}
when $p_{\phi}(z|x)$ is the same with $p(z)$. In other words, this posterior collapse property of VAEs suggests that, the generator is encoder-agnostic if the encoder can perfectly model $p(z)$. Minimising the KL divergence between $p_{\phi}(z|x)$ and $p(z)$ during training VAEs actually decreases the dependency between VAE generators and VAE encoders.However, in practice, the KL divergence should not be reduced to 0. Otherwise, the VAE falls into a posterior collapse problem. That is, the VAE becomes inexpressive and cannot reconstruct data examples. Therefore, in practice, the MVD should estimate the mean and the variance of the encoded distributions. Otherwise, the MVD cannot correctly decode the output from black-box encoders.

To tackle this problem, we first calibrate the intake distribution of the MVD to approximate the encoded distribution from the attacked black-box VAE encoder. We query the attacked encoder using $X$. We then pair the query outputs $Z$ with $X$ to form the training set $(z, x), z\in Z, x\in X$. Therefore, we can calibrate the MVD by training it to map $z$ back to $x$. We consider that the calibration is feasible on the attacker side since the attacker launches the attack as an insider. However, the minimal number of queries required for the calibration is also investigated in the evaluation section. After the calibration, the trained MVD can take inputs from they VAE encoder and then generate semantically correct adversarial examples. Upon this stage, the attacker can launch attacks in both the double-black-box setting and the black-box-encoder setting.

\subsection{Adversarial objective and optimisation}\label{S::Method::2}
Given an input image $x$, a benign encoder $e$, and a victim classifier $f$, the aim of the MVD is to find the parameters $\theta$ that minimise an adversarial loss within a distortion budget $\epsilon$. 
\begin{align}\label{eq2}
\theta = &\argmin L_{adv}\{f(g_{\theta}(e(x))), f(e(x))\}\\
&s.t.\ \|g_{\theta}(x) - x\| \leq \epsilon. \nonumber
\end{align}

In Equation \ref{eq2}, $L_{adv}$ is a loss function which sets the adversarial goal to be optimised towards. Optimising Equation \ref{eq1} is difficult. Therefore, we convert the optimisation problem to its relaxed form as follows:
\begin{equation}\label{eq3}
\theta = \argmin c\cdot L_{adv}\{f(g_{\theta}(e(x))), f(e(x))\} + \|g_{\theta}(x) - x\|_p.
\end{equation}

In the objective function, $c$ is a constant that balances the adversarial loss and the $p$-norm of the perturbation. In order to make the objective function differentiable for the later optimisation, we set $p$ to 2 here. We set the adversarial loss function $L_{adv}$ as follows:
\begin{equation}\label{eq4}
L_{adv}(\theta) = max(max\{Z_{\theta}^{i\neq t}[e(x)]\}-Z_{\theta}^{t}[e(x)], -\kappa),
\end{equation}
wherein $Z$ stands for the output logits of the classifier. $t$ is the target class of misclassification. $\kappa$ is a parameter controlling the attack strength. The MVD trained under higher $\kappa$ will produce adversarial examples with higher confidence of misclassification.

\subsection{Framework of MVD-based MitM attacks}\label{S::Method::3}
We divide the framework of MVD-based MitM attacks into three phases. In the first phase, we first calibrate the latent distribution of the MVD. Specifically, we concatenate the untrained MVD to a VAE encoder to form an untrained VAE. The VAE is trained following the training subroutine introduced in Section \ref{S:vae} to estimate target VAEs. Upon trained, the trained VAE contains the encoder $p_{\phi}(x|z)$ parameterised by $\phi$ and the MVD $q_{\theta}(x|z)$ parameterised by $\theta$. The latent distribution of $q_{\theta}(x|z)$ is now calibrated to the vicinity of a standard normal distribution. Next, we turn the MVD into an adversarial example generator. We freeze the $\phi$ of the encoder and incrementally train the $\theta$ to minimise Equation (\ref{eq3}). We use Adam as the optimiser during the first and the second phase. Compared to attacks such as advGAN, the two-step training of MVD enables rapid modification of adversarial objectives (\textit{e.g.}, switching between non-targeted attack and targeted attacks).

In the third phase, the attacker can either build an adversarial VAE between raw inputs and a classifier or insert the MVD between an arbitrary VAE encoder and a victim classifier to launch attacks. In this framework, the dimensionality of the inputs/outputs of the VAE encoder/victim classifier should be obtained by the attacker prior to launching the attacks. Consider that the input/output information of DNN models is commonly released to users, this prerequisite is painless to fulfilled. As a conclusion, the framework of building the malicious decoder is illustrated in Fig \ref{framework}. We also outline the critical steps of training MVD in Algorithm \ref{algo1}. In the algorithm, $L$ is log loss.

\begin{algorithm}[h!]
\footnotesize
\caption{MVD attacks}
\label{algo1}
\textbf{Phase 1}: \\
\textbf{Input}: $X$, $f$, $vae\_iter$, $adv\_iter$\\
\textbf{Initialisation}: $\theta$, $\phi$ \\
\For{$i$ in $vae\_iter$}
	{
	$\theta=\theta-\nabla_{\theta}E_{x\in X}L(x, g_{\theta}(e_{\phi}(x)))$\\
	$\phi=\phi-\nabla_{\phi}E_{x\in X}[KL(e_{\phi}(x)||\mathcal{N}(0,1))+L(x, g_{\theta}(e_{\phi}(x)))]$\\	
	}
\If{$e$ is a black box}
{ 
	\For{$i$ in $vae\_iter$}
		{
		$\theta=\theta-\nabla_{\theta}E_{x\in X}L(x, g_{\theta}(e(x)))$\\
		}
}
\textbf{Phase 2}: \\
\For{$i$ in $adv\_iter$}
    {	
  	$\theta=\theta-\nabla_{\theta}E_{x\in X}\{c\cdot L_{adv}[f(g_{\theta}(e(x))),f(e(x))] + \|g_{\theta}(x)-x\|_p\}$\\
    }
\textbf{Output}: $\theta$.

\textbf{Phase 3}: \\
\textbf{Input}: $\theta$, $e(x)$, $f$ \\
\textbf{Output}: $f(g_{\theta}(e(x)))$\\
\end{algorithm}

\section{Evaluation}\label{S::Eval} 
We present the evaluation of the MVD attack in this section. We first summarise the experimental settings for the evaluation. Subsequently, we present the evaluation results of the MVD attacks in the double-black-box setting, the black-box-classifier setting, the black-box-encoder setting, and the double-white-box setting. In the evaluation, we compared the MVD attack with another three state-of-the-art attacks, namely Fast Gradient Sign (FGSM)\footnote{\textit{https://github.com/tensorflow/cleverhans}} \cite{Goodfellow2014Explaining}, Calini$\&$Wagner\footnote{\textit{https://github.com/tensorflow/cleverhans}} (C$\&$W) \cite{carlini2017adversarial}, and advGAN\footnote{\textit{https://github.com/ctargon/AdvGAN-tf}} \cite{xiao2018generating}. FGSM and C$\&$W are typical query-and-optimised attacks, while advGAN is an adversarial example generator that can attack on-the-fly. We also compared the MVD with the three attacks on the computing overhead and the magnitude of the adversarial perturbations.

\subsection{Experiment setups}
We included MNIST dataset and CIFAR10 dataset in the evaluation. For each dataset, we evaluated the performance of the evaluated attacks in both black-box settings and white-box settings. For the sake of comparison, we included the double-black-box, black-box-classifier, and black-box-encoder settings in the black-box evaluation of MVD attacks. All experiments were running on a server with an RTX-20180-ti GPU and 128G memory. We adopted the original training-testing data split from MNIST and CIFAR10 during training and testing the MVD. We adopted two different MVD architectures to cope with the two datasets. 

In the black-box setting, the MVD cannot access the parameters of both the victim classifier and the victim encoder. Therefore, we first train a shadow classifier to approximate the victim. We then train the MVD based on the trained shadow classifier. In the white-box setting, we directly train the MVD based on the victim classifier. All the classifiers achieved accuracy above $95\%$ and $80\%$ on the MNIST examples and the CIFAR10 examples, respectively. The architectures of the adopted classifiers are summarised in Table \ref{classifier_arc}. The architectures of the VAE encoders are in Table \ref{vae_arc}. The structures of the MVD is presented in Table \ref{mvd_arc}. We ran a grid search from $\{0.003, 0.01, 0.05, 0.1, 1\}$ to select the best constant $c$ in the adversarial loss function \ref{eq4}. For other attacks, we used the recommend hyper-parameters from the original papers. As for advGAN, we did not adopt the static/dynamic distillation process in our evaluation since we wanted to set the attacks under the same prerequisites. For the reference, the hyper-parameter settings of FGSM, C$\&$W, advGAN, and MVD during the comparisons are summarised in Table \ref{attack_params}. In the table, the hyper-parameters with a superscript $m$ were used in the evaluation on the MNIST dataset, and those with a superscript $c$ were used for the CIFAR10 dataset.

\begin{table}
\caption{Classifier Architectures}
\label{classifier_arc}
\centering
\resizebox{\columnwidth}{!}{%
\begin{tabular}{c c c c}
\hline
Shadow(MNIST) & Victim(MNIST) & Shadow(CIFAR10) & Victim(CIFAR10)\\
\hline
Input 28$\times$28$\times$1 & Input 28$\times$28$\times$1 & Input 32$\times$32$\times$3 & Input 32$\times$32$\times$3\\
Convo 32$\times$3$\times$3 & Convo 32$\times$3$\times$3 & Convo 32$\times$3$\times$3 & Convo 32$\times$3$\times$3\\
Convo 32$\times$3$\times$3 & Convo 32$\times$3$\times$3 & Convo 32$\times$3$\times$3 & Convo 32$\times$3$\times$3\\
MaxPooling 2$\times$2 & Dropout 0.2 & MaxPooling 2$\times$2 & MaxPooling 2$\times$2 \\
Convo 64$\times$3$\times$3 & FC 128 & Dropout 0.2 & Dropout 0.2\\
Convo 64$\times$3$\times$3 & Dropout 0.2 & Convo 64$\times$3$\times$3 & Convo 64$\times$3$\times$3\\
MaxPooling 2$\times$2 & Softmax FC 10 & Convo 64$\times$3$\times$3 & Convo 64$\times$3$\times$3\\
FC 200 & - & MaxPooling 2$\times$2 & MaxPooling 2$\times$2\\
FC 200 & - & Dropout 0.2 & Dropout 0.2\\
Softmax FC 10 & - & Convo 128$\times$3$\times$3 & Convo 128$\times$3$\times$3\\
- & - & Convo 128$\times$3$\times$3 & Convo 128$\times$3$\times$3\\
- & - & MaxPooling 2$\times$2 & MaxPooling 2$\times$2\\
- & - & Dropout 0.2 & Dropout 0.2\\
- & - & FC 512 & FC 200\\
- & - & Dropout 0.2 & Softmax FC 10\\
- & - & Softmax FC 10 & - \\
\hline
\end{tabular}%
}
\end{table}

\begin{table}
\caption{VAE Encoder Architectures}
\label{vae_arc}
\centering
\resizebox{\columnwidth}{!}{%
\begin{tabular}{c c c c}
\hline
Shadow(MNIST) & Victim(MNIST) & Shadow(CIFAR10) & Victim(CIFAR10)\\
\hline
Input 784 & Input 784 & Input 32$\times$32$\times$3 & Input 32$\times$32$\times$3\\
FC 512 & FC 600 & Convo 3$\times$2$\times$1 & Convo 3$\times$2$\times$1\\
FC 10 & FC 10 & Convo 32$\times$2$\times$2 & Convo 32$\times$2$\times$2\\
- & - & Convo 32$\times$3$\times$1 & Convo 32$\times$3$\times$1\\
- & - & Convo 32$\times$3$\times$1 & Convo 32$\times$3$\times$1\\
- & - & Flatten & Flatten\\
- & - & FC 512 & FC 1024\\
- & - & FC 128 & FC 128\\
\hline
\end{tabular}%
}
\end{table}

\begin{table}
\caption{MVD Architectures}
\label{mvd_arc}
\centering
\begin{tabular}{c c}
\hline
MVD(MNIST)&MVD(CIFAR10)\\
\hline
FC 10 & FC 128\\
FC 512 & FC 512\\
FC 784 & FC 8192 \\
- & Reshape 16$\times$16$\times$32 \\
- & Deconvo 32$\times$3$\times$1 \\
- & Deconvo 32$\times$3$\times$1 \\
- & Deconvo 32$\times$2$\times$2 \\
- & Deconvo 3$\times$1$\times$1 \\
\hline
\end{tabular}
\end{table}

\begin{table}
\caption{Hyper-parameters of the Attacks}
\label{attack_params}
\centering
\begin{tabular}{c c}
\hline
Attack & Hyper-Parameter\\
\hline
FGSM & $\eta=\{0.3^m, 0.1^c\}$\\
C$\&$W & $\kappa=0^{m,c}$\\
advGAN & $c=\{0.3^m, 8^c\}$\\
MVD & $c\in\{0.01^m, 0.003^c\}$, $\kappa=0^{m,c}$\\
\hline
\end{tabular}
\end{table}

We selected 1,000 MNIST examples and 1,000 CIFAR10 examples as our evaluation datasets. For making a fair comparison, we ablated the errors brought by the VAEs and the classifiers during the evaluation. Specifically, the selected examples in the evaluation dataset held the following two attributes: 1) the examples can be correctly classified by the classifier in individual evaluations, and 2) the classifier can correctly classify their reconstructed counterparts from the benign VAEs in individual evaluations. Otherwise, the evaluation datasets were randomly drawn from the original MNIST/CIFAR10 test dataset.

\subsection{Attacks in black-box settings}\label{blackbox_eval}
\begin{figure}[t!]
\center
\includegraphics[width=1\linewidth]{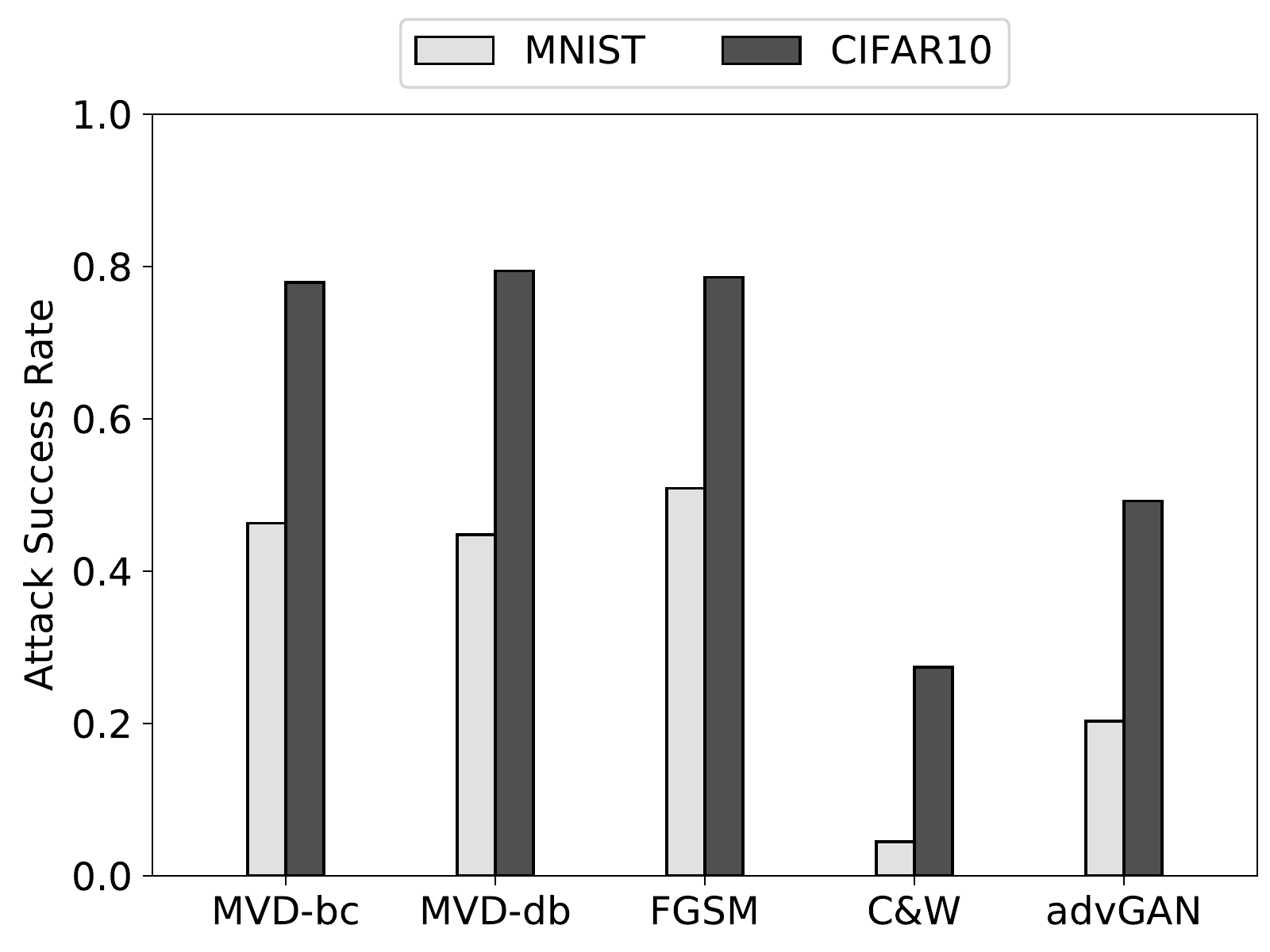}
\caption{Attack success rates of the non-targeted attacks in the black-box setting.}
\label{bb_suc}
\end{figure}

\begin{figure}[t!]
\center
\includegraphics[width=1\linewidth]{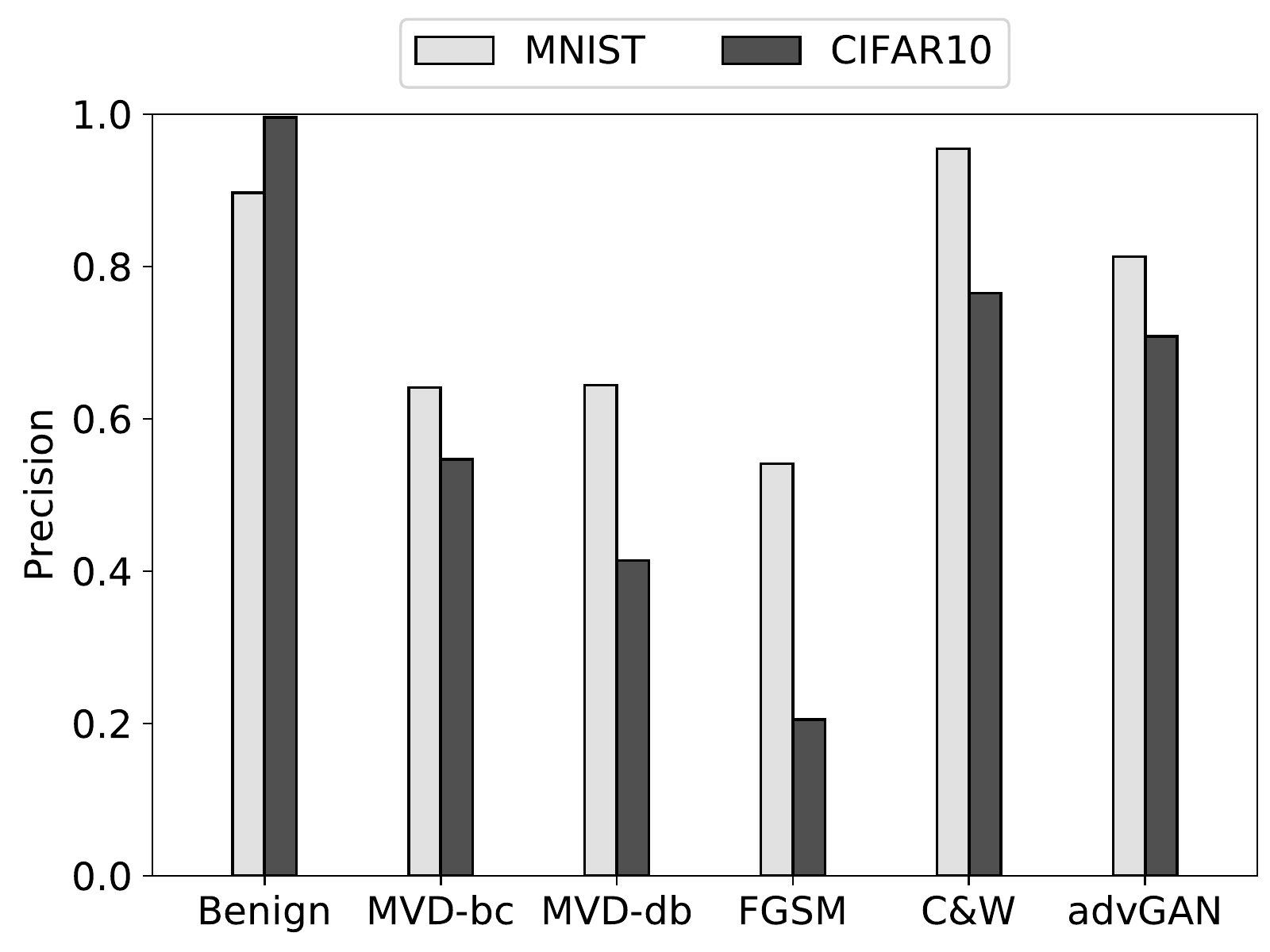}
\caption{Precision scores of the victim classifier against the non-targeted attacks in the black-box setting.}
\label{bb_precision}
\end{figure}

\begin{figure}[t!]
\center
\includegraphics[width=1\linewidth]{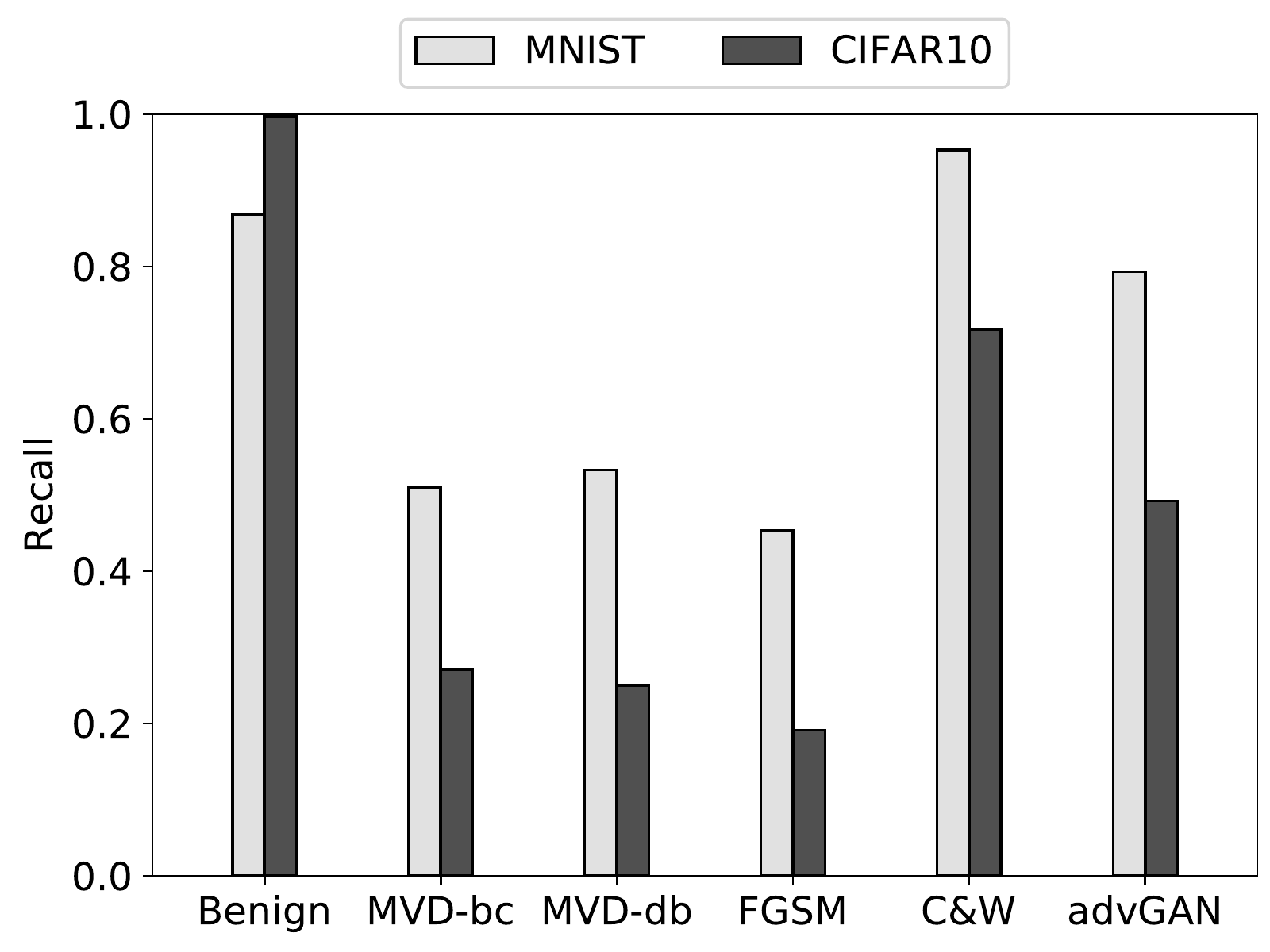}
\caption{Recall scores of the victim classifier against the non-targeted attacks in the black-box setting.}
\label{bb_recall}
\end{figure}

\begin{figure}[t!]
\center
\includegraphics[width=1\linewidth]{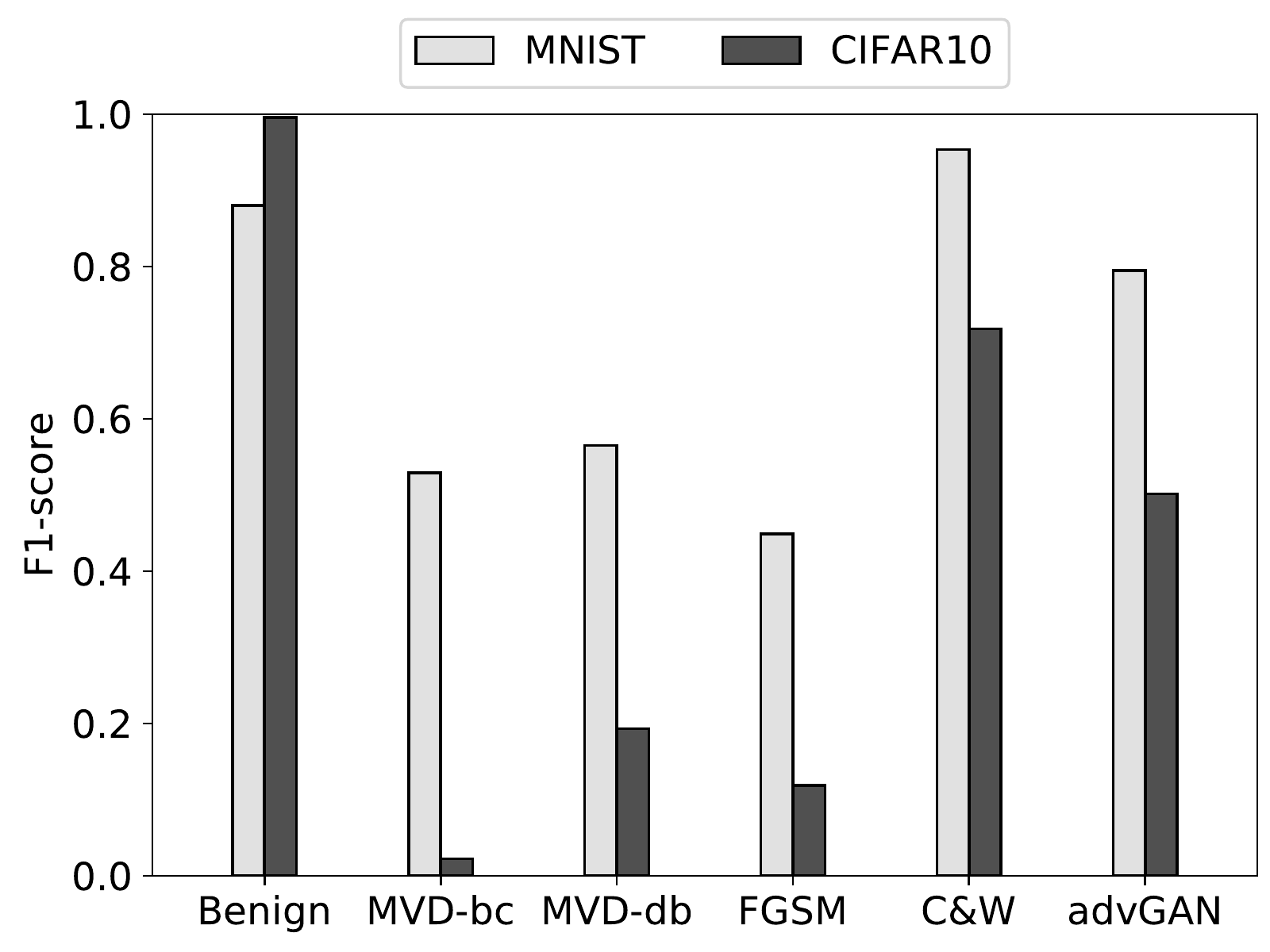}
\caption{F1-score scores of the victim classifier  against the non-targeted attacks in the black-box setting.}
\label{bb_fscore}
\end{figure}

We first evaluated the MitM attack in black-box settings. We evaluated MVD attacks in the double-black-box and black-box-classifier settings. For FGSM, C$\&$W, and advGAN, we considered that the classifier is a black-box. Hence, in the MVD attacks, we first trained a shadow VAE composed of a shadow VAE encoder and an MVD. We also trained another VAE and used its encoder as the black-box encoder. Next, we employed a shadow classifier trained to estimate the victim model into the training process. The trained shadow VAE was then concatenated to either the shadow classifier for training the MVD to be malicious. We used the trained MVDs to craft adversarial examples of both MNIST and CIFAR10 data. Regarding to other attacks, we adopted them to craft adversarial examples based on the same shadow classifier. In the attacking stage, we applied the MVDs and the other attacks to attack the same victim classifier. We compared the success rate (\textit{i.e.}, ratio of the attacks that lead to misclassification) of the attacks. We also compared the precision, recall, and F1-score of the victim classifier under attacks from the MVDs with that of the same classifier attacked by FGSM, C$\&$W, and advGAN. The comparisons of the attacks are shown in Figure \ref{bb_suc}, \ref{bb_precision}, \ref{bb_recall}, and \ref{bb_fscore}. In this stage, all the attacks were non-targeted (\textit{i.e.}, without specifying a misclassification target).

In the plots, MVD-bc notates the MVD in the black-box-classifier setting, and MVD-db stands for the MVD in the black-box-classifier setting. According to the plotted results, we could observe that the MVD attacks outperformed C$\&$W and advGAN in terms of attack success rate, on both MNIST and CIFAR10 datasets. As for C$\&$W attacks, it may increase the misclassification confidence to enhance its success rate. However, the MVD attacks could also increase its misclassification confidence. Also, there is a correlation between the misclassification confidence and the perceptibility of the C$\&$W perturbations. Thus, here we only made the comparison by applying a border misclassification confidence (\textit{i.e.}, $\kappa=0$) to both the C$\&$W and the MVD. Compared to the performance on benign examples, the precision, recall, and F1-score of the victim classifier were significantly reduced by the MVD attacks. In most cases, compared to the other attacks, the MVD attacks induced more loss of the classifier performance.

\begin{figure}[t!]
\center
\includegraphics[width=1\linewidth]{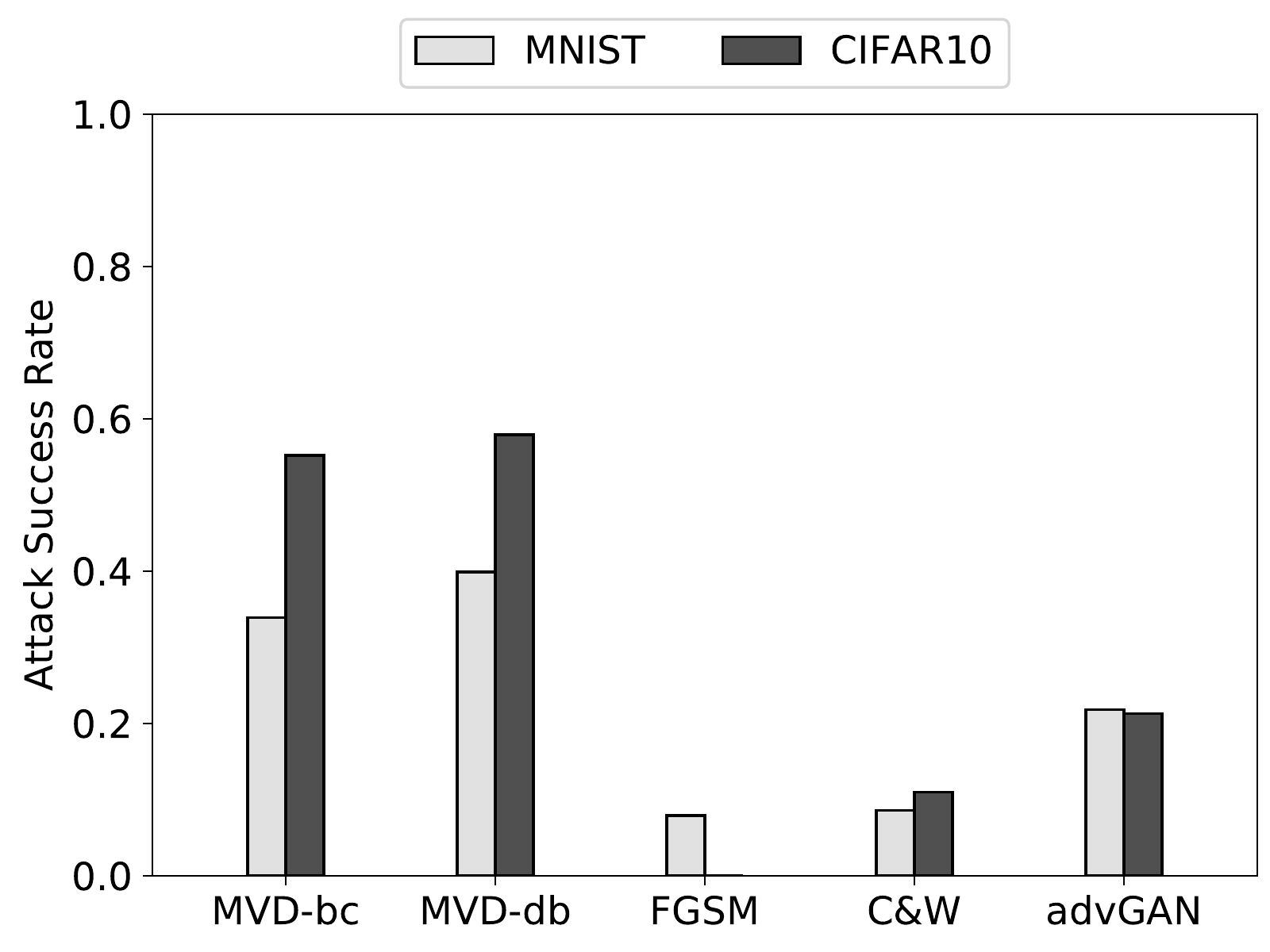}
\caption{Attack success rates of the attacks targeting at a random class, in the black-box settings.}
\label{bb_suc_t}
\end{figure}

To better understand the characteristics of the MVD during targeted attacks, we trained ten MVDs, each of which is trained to produce a unique misclassification target selected from 0 to 9. We repeated this process for all the black-box settings (\textit{i.e.}, black-box-classifier and double-black-box settings). We recorded the attack success rates of the MVDs on the victim classifier in Table \ref{suc_targeted}. We again adopted MVD, FGSM, C$\&$W, and advGAN to craft 1000 each adversarial examples that target at a random class. We compared the attack success rates of the four attacks in Figure \ref{bb_suc_t}. Herein, MVD-be stands for the MVD in the black-box-encoder setting, and MVD-dw indicates that the MVD is in the double-white-box setting. 

\begin{table*}
\caption{Success Rate of Targeted Attacks}
\label{suc_targeted}
\centering
\begin{tabular}{c|ccc|ccc}
\hline
\multirow{2}{*}{Threat Model} & \multicolumn{3}{c|}{MNIST} & \multicolumn{3}{c}{CIFAR10} \\ \cline{2-7} 
                              & Worst  & Average  & Best   & Worst   & Average   & Best   \\ \hline
Black-box-classifier Setting  & 0.178  & 0.382    & 0.615  & 0.311   & 0.5861    & 0.9    \\
Double-black-box Setting      & 0.157  & 0.376    & 0.564  & 0.248   & 0.581     & 0.9    \\
Black-box-encoder Setting     & 0.957  & 0.979    & 0.993  & 0.987   & 0.997     & 1      \\
Double-white-box Setting      & 0.961  & 0.976    & 0.992  & 0.99    & 0.996     & 1      \\ \hline
\end{tabular}
\end{table*}

According to Table \ref{suc_targeted}, when the victim classifier was a black-box, the MVD achieved over 90$\%$ success rate with the CIFAR10 adversarial examples in the best case. The MNIST examples had over 60$\%$ success rate. Since the error brought by the VAE was ablated from the experiment, the results suggested that the attacks can transfer better among the shadow classifier and victim classifier for CIFAR10. Importantly, we could observe that the MVD achieved similar success rate in both the black-box-classifier and double-black-box setting, which suggests that the MVD well decoded the outputs from the black-box encoder.

\subsection{Attacks in white-box settings}\label{whitebox_eval}
\begin{figure}[t!]
\center
\includegraphics[width=1\linewidth]{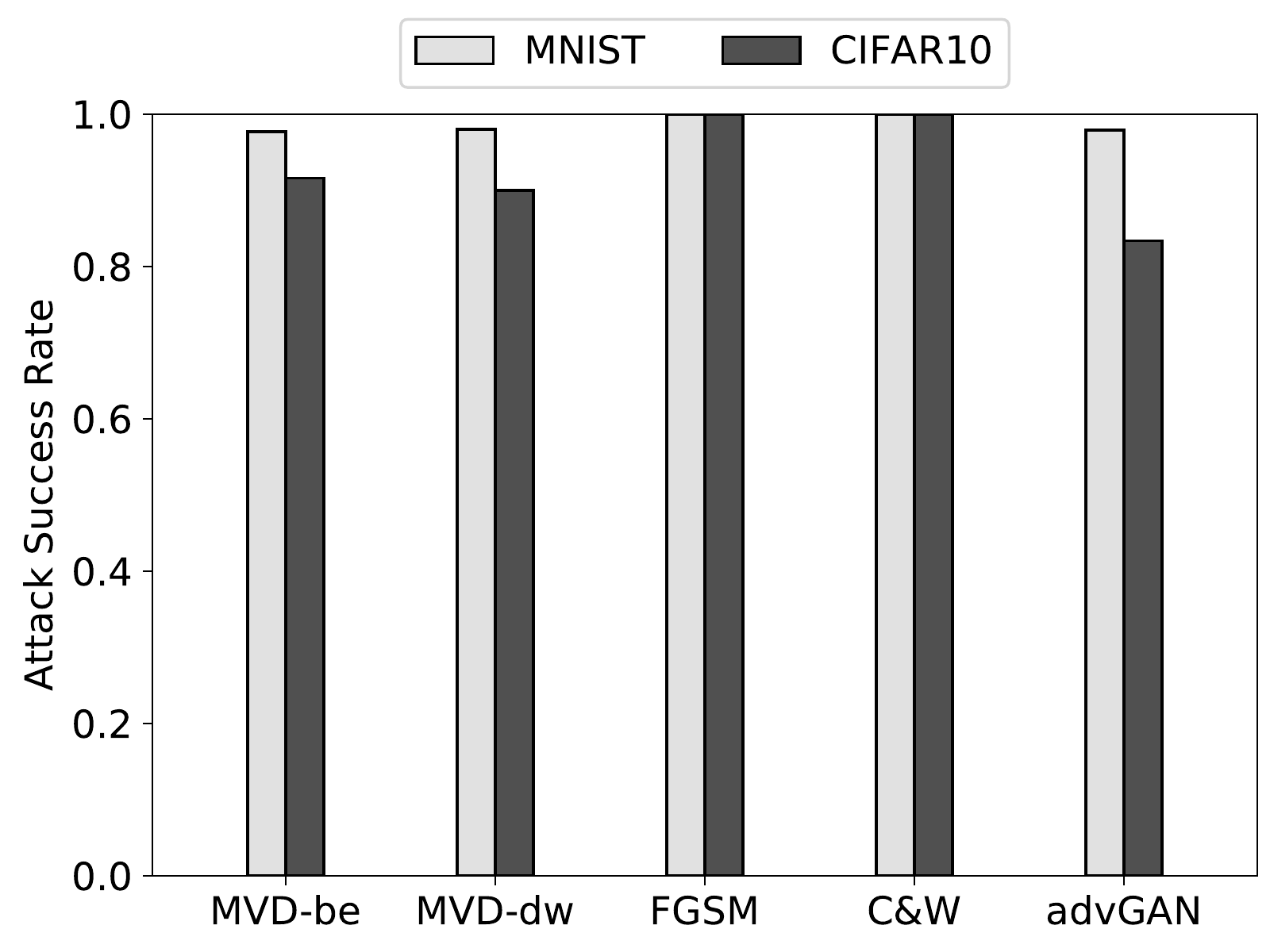}
\caption{Attack success rates of the attacks in the white-box setting.}
\label{wb_acc}
\end{figure}

\begin{figure}[t!]
\center
\includegraphics[width=1\linewidth]{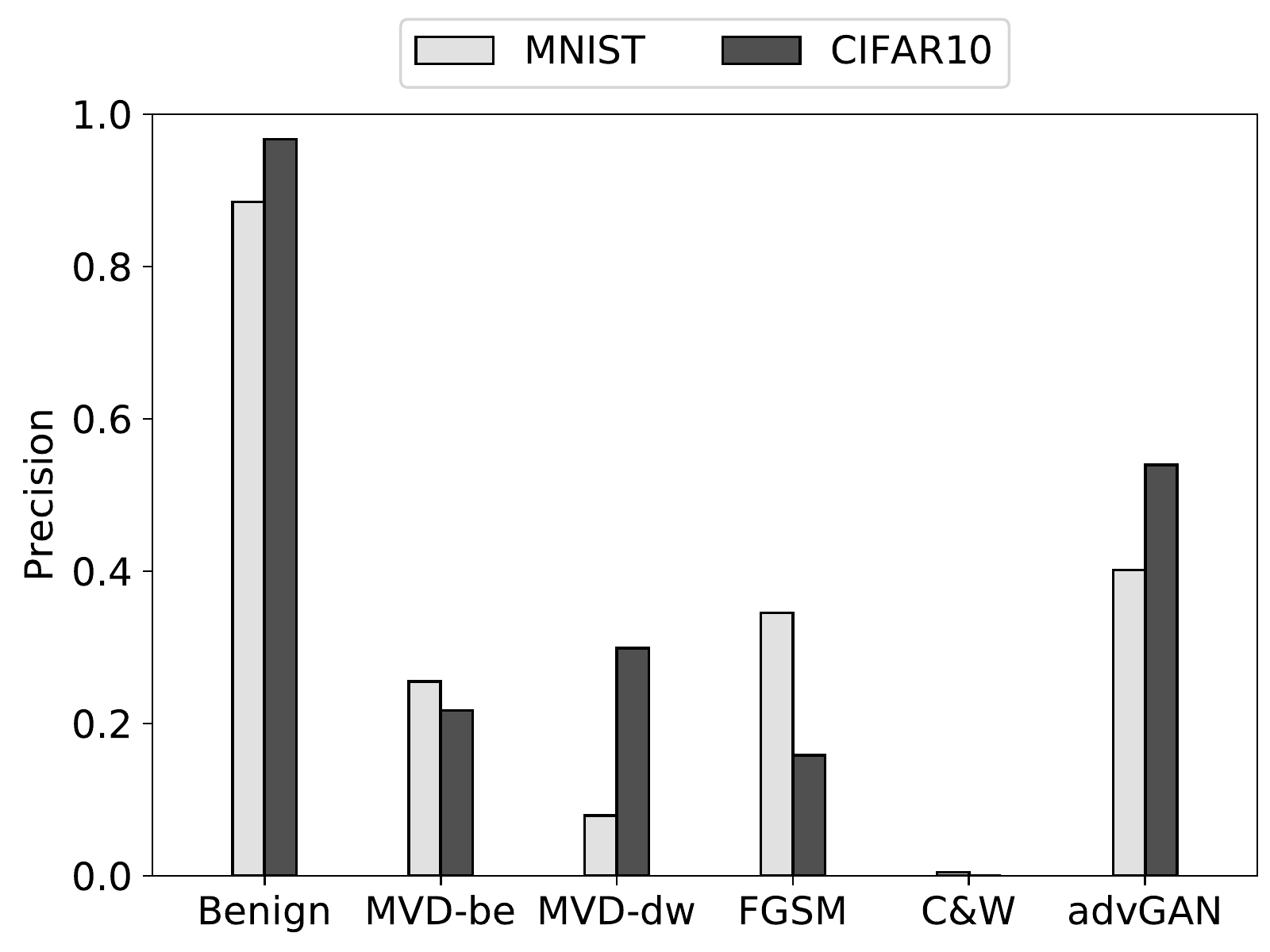}
\caption{Precision scores of the victim classifier in the white-box setting.}
\label{wb_precision}
\end{figure}

\begin{figure}[t!]
\center
\includegraphics[width=1\linewidth]{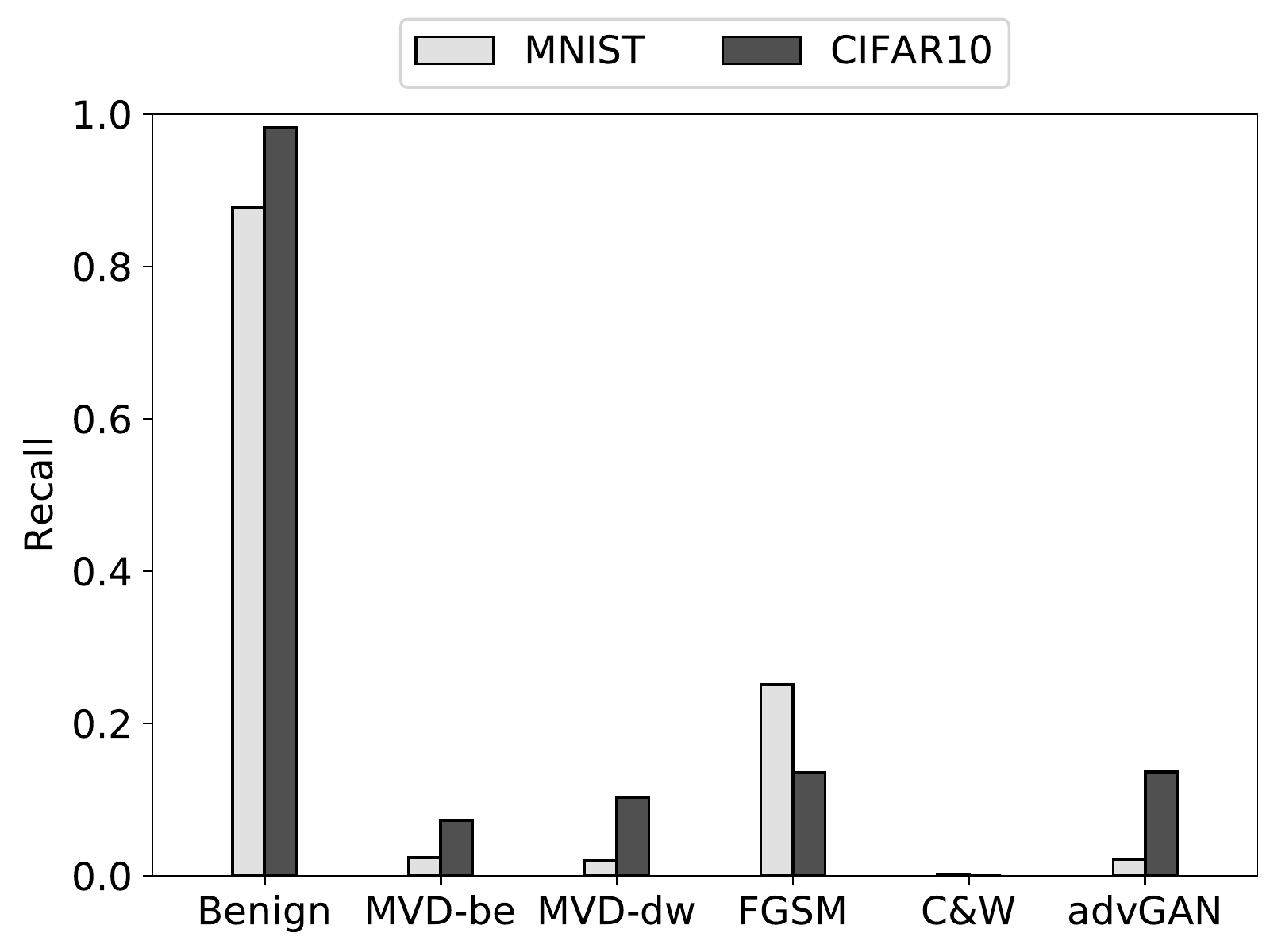}
\caption{Recall scores of the victim classifier in the white-box setting.}
\label{wb_recall}
\end{figure}

\begin{figure}[t!]
\center
\includegraphics[width=1\linewidth]{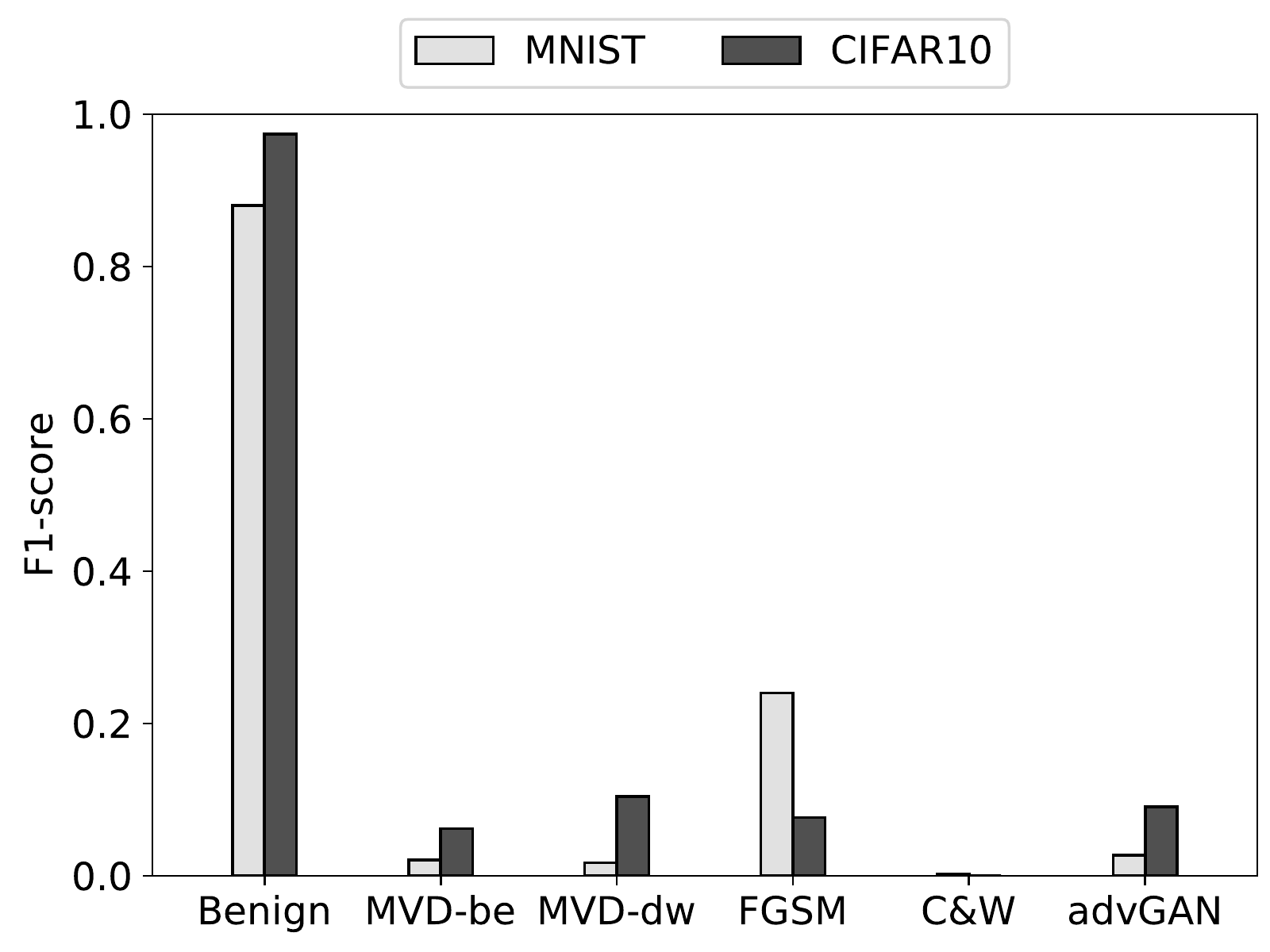}
\caption{F1-score scores of the victim classifier in the white-box setting.}
\label{wb_fscore}
\end{figure}

We evaluated the performance of the victim classifier attacked in white-box settings. As for MVD attacks, we evaluate it in the and black-box-encoder setting and the double-white-box setting. Particularly, the shadow VAE encoder was regarded as a white-box victim encoder in this setting. Regarding all the evaluated attacks, the attackers can access the parameters of the victim classifier. In this setting, MVD and advGAN were trained based on the first-order information of the victim classifier. We concatenated the MVD onto the victim classifier and trained the MVD based on the method depicted in Fig \ref{framework}. FGSM and C$\&$W attacks directly generated adversarial examples based on the gradients of the victim classifier. Given the test dataset from MNIST/CIFAR10, we recorded the accuracy, precision, recall, and F1-score of the victim classifier under the white-box attacks. The comparisons are shown in Figure \ref{wb_acc}, \ref{wb_precision}, \ref{wb_recall}, and \ref{wb_fscore}. According to the results, the precision, recall, and F1-score of the classifier were significantly reduced by the MVD-be and MVD-dw attacks.

\begin{figure}[t!]
\center
\includegraphics[width=1\linewidth]{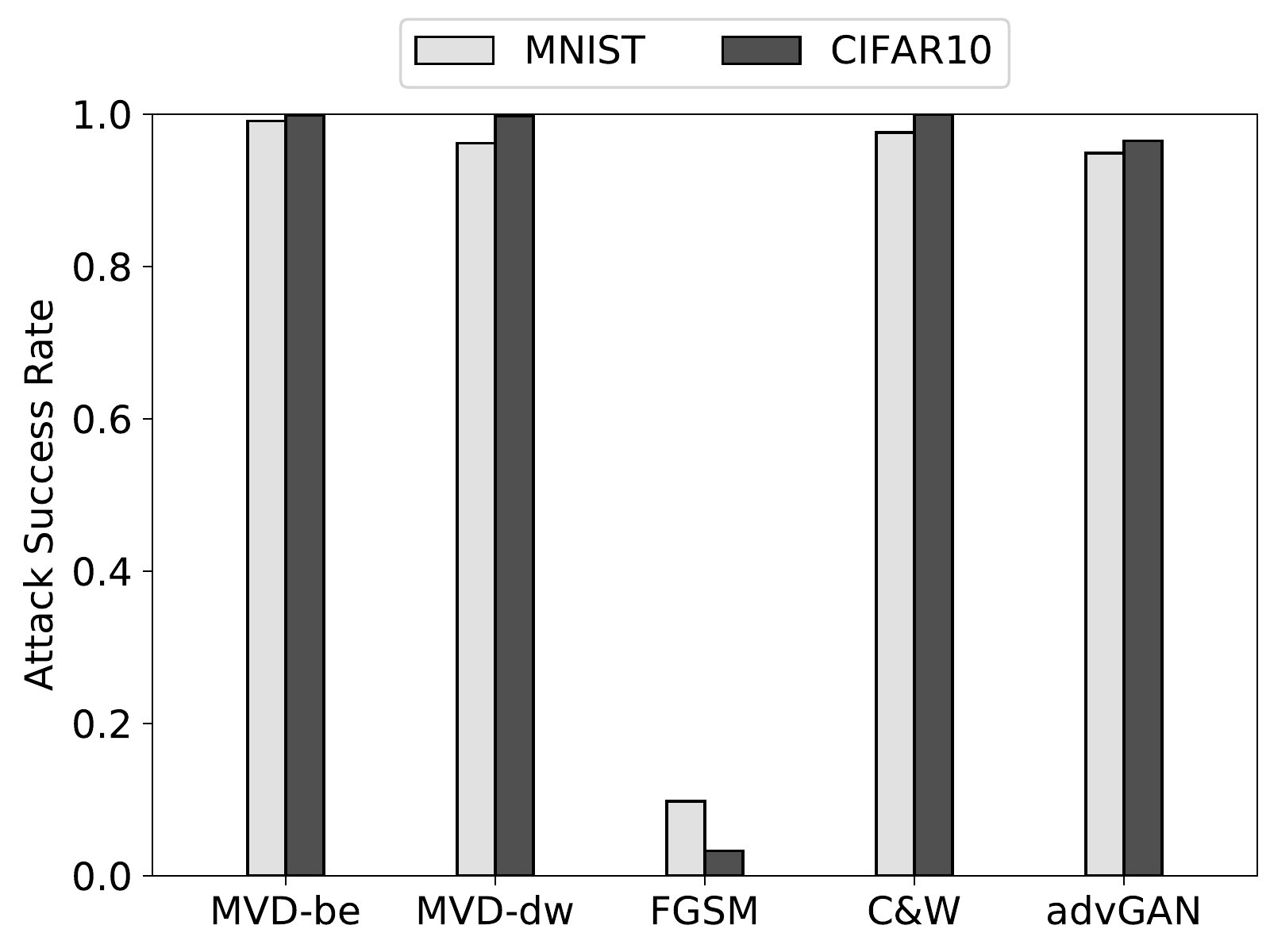}
\caption{Attack success rates of the attacks targeting at a random class, in the white-box setting.}
\label{wb_suc_t}
\end{figure}

Similar to that in the black-box setting evaluation, we trained 10 MVD targeting at different data categories, in the black-box-encoder setting and the white-box setting, respectively. We then recorded the minimum, mean, and maximum of the MVD attack success rates against the victim classifier in Table \ref{suc_targeted}. In both the black-box-encoder and double-white-box settings, the MVD attacks achieved above 95$\%$ minimal success rate. Moreover, the black-box encoder barely affected the attack success rate. We also compared the performance of all the four attacks on random misclassification targets. We plotted their success rates in Figure \ref{wb_suc_t}. It could be observed that, on both MNIST and CIFAR10 datasets, the MVD targeted attacks achieved equivalent success rate with C$\&$W and advGAN.

\subsection{Overhead and distortion}
In this section, we compared the runtime of MVD with that of FGSM, C$\&$W, and advGAN. We used the four methods to generate 1000 MNIST/CIFAR10 adversarial examples against the victim classifier, which was a white-box. For each attack, we averaged the recorded runtime over the 1000 examples to calculate the average runtime for generating one adversarial example. We summarise the averaged runtime in Table \ref{overhead}. It can be observed that the MVD has achieved comparable overhead with that of advGAN. The runtime of MVD is about $10\to30$ times less than that of advGAN, $10^{4}$ times less than that of FGSM, and more than $3\times 10^{5}$ times less than that of C$\&$W. 

\begin{figure*}
\centering
\begin{subfigure}{0.5\textwidth}
  \centering
  \includegraphics[width=1\linewidth]{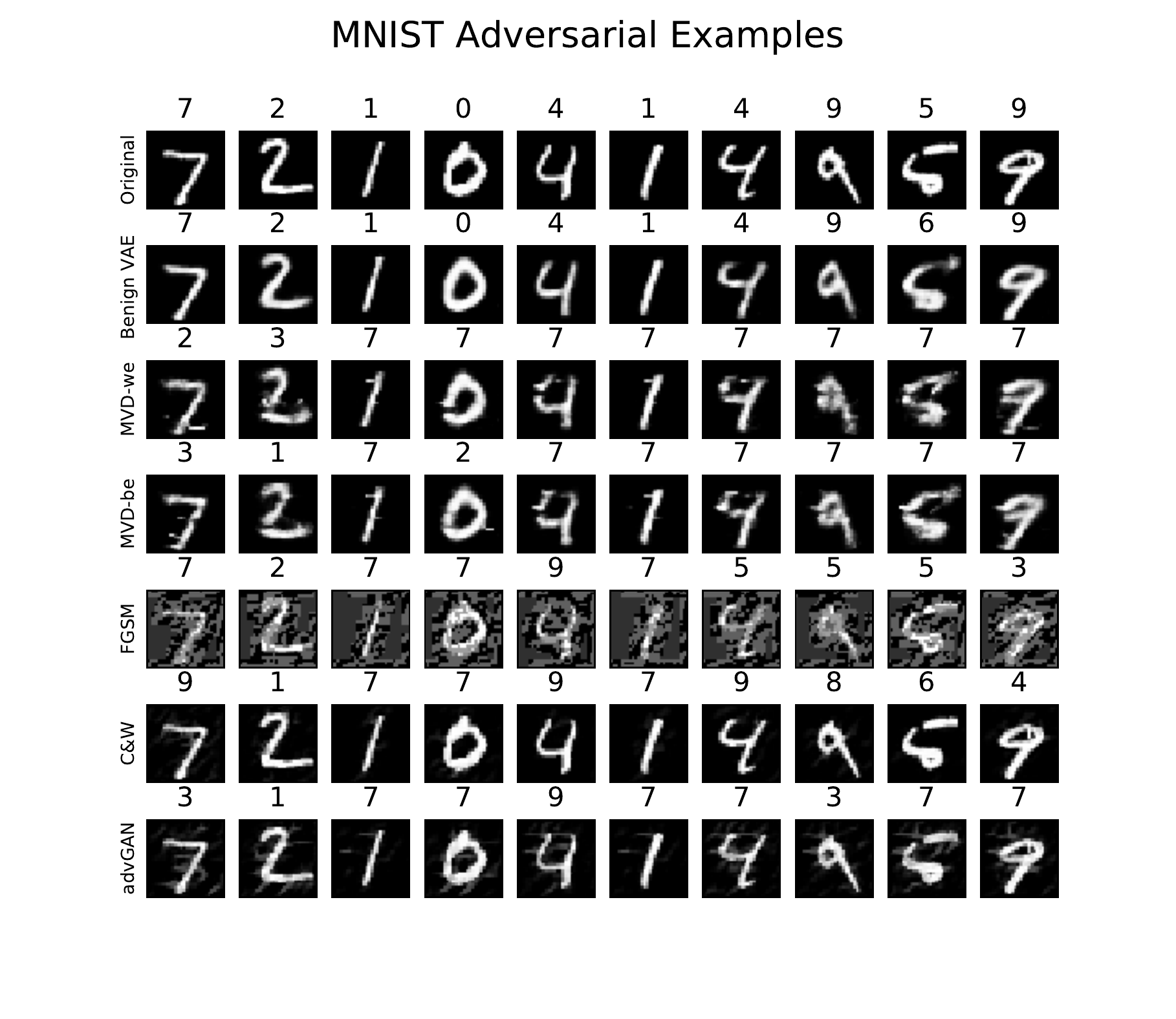}
  \label{mnist_demo}
\end{subfigure}%
\begin{subfigure}{0.525\textwidth}
  \centering
  \includegraphics[width=1\linewidth]{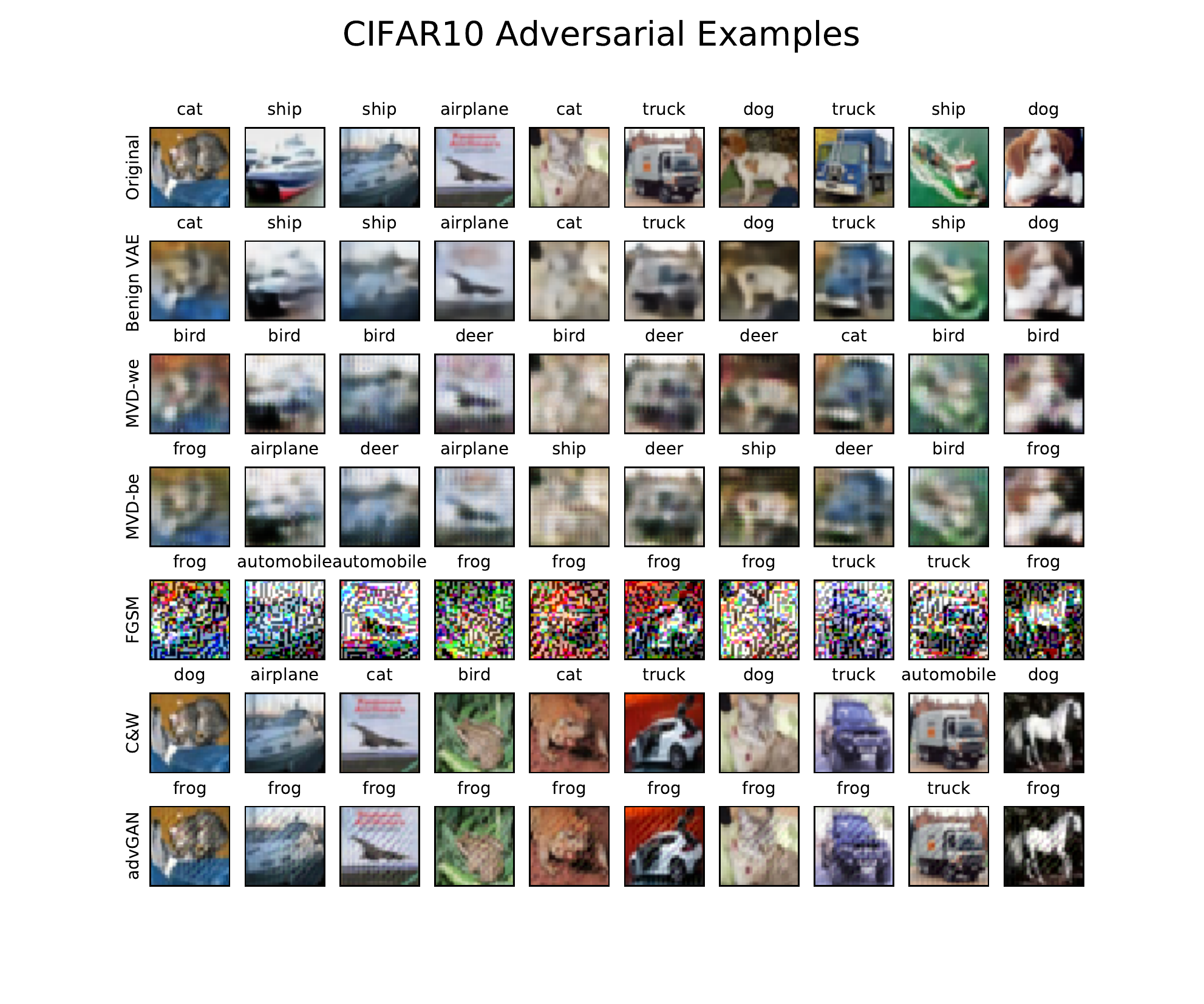}
  \label{cifar10_demo}
\end{subfigure}
\caption{Demo adversarial examples generated by MVD and other attacks against the same white-box classifier (no cherry-picking). In the first row are the original examples. Their counterparts reconstructed by the benign victim VAE are in the second row. MVD-we indicates that the VAE encoder is a white-box for the MVD. MVD-be stands for MVD in the black-box-encoder setting. The number above each example is its classification results from the classifier. The MVD perturbations of the MNIST examples had an acceptable visual perceptibility. The CIFAR10 examples are blurry due to the VAE reconstruction process. However, the perturbation added to the reconstructed examples are acceptable.}
\label{demo}
\end{figure*}

Next, we computed the minimum, mean, and maximum of the $L_2$ norms of the perturbations. The results are in Table \ref{dist}. According to the comparison, MVD resulted in comparable distortions with that of advGAN. We also randomly selected ten examples from each dataset and visualised their adversarial examples generated by different attacking methods, in Figure \ref{demo}. The CIFAR10 examples generated by the MVD are blurry compared to that of the other attacks. However, this was due to the reconstruction quality of the benign VAEs. Techniques such as VAE-GAN\cite{larsen2016autoencoding} or Two-stage VAE\cite{dai2019diagnosing} can improve the fidelity of the VAEs.  On the MNIST examples, the adversarial perturbations generated by the MVD attacks are comparable to that of advGAN. Interestingly, instead of masking perturbations onto the examples as the other attacks did, the MVD decoded the examples to be adversarial. As a result, the MVD attacks slight transformed the image content without changing the visual-semantics of the examples.

\begin{table}
\caption{Runtime Comparison}
\label{overhead}
\centering
\begin{tabular}{c|c c c c}
\hline
& MVD & FGSM & C$\&$W & advGAN\\
\hline
MNIST & $2\times 10^{-4}$s & $1.88$s & $7.75\times 10^1$s & $1.85\times 10^{-3}$s\\
CIFAR10 & $3.5\times 10^{-4}$s & $1.97$s & $5.76\times 10^1$s & $1.05\times 10^{-2}$s\\
\hline
\end{tabular}
\end{table}

\begin{table*}
\caption{Distortion Comparison}
\label{dist}
\centering
\begin{tabular}{c|c|c|c|c|c|c|c|c|cc}
\hline
        & \multicolumn{2}{c|}{MVD-be} & \multicolumn{2}{c|}{MVD-we} & \multicolumn{2}{c|}{FGSM} & \multicolumn{2}{c|}{C\&W} & \multicolumn{2}{c}{advGAN}           \\ \hline
        & MNIST       & CIFAR10       & MNIST       & CIFAR10       & MNIST      & CIFAR10      & MNIST      & CIFAR10      & \multicolumn{1}{c|}{MNIST} & CIFAR10 \\ \hline
Minimum & 1.694       & 2.319         & 1.697       & 2.278         & 5.917      & 14.586       & 0.003      & 0.003        & \multicolumn{1}{c|}{0.862} & 1.197   \\
Mean    & 3.925       & 4.802         & 3.986       & 4.783         & 7.454      & 16.601       & 0.909      & 0.158        & \multicolumn{1}{c|}{2.207} & 1.836   \\
Maximum & 6.88        & 7.276         & 6.719       & 7.531         & 8.238      & 16.627       & 2.152      & 0.626        & \multicolumn{1}{c|}{3.646} & 2.913   \\ \hline
\end{tabular}
\end{table*}

\subsection{Diagnosing the transferability of MVD attacks}
The transferability of adversarial examples affects the attack success rate in black-box settings. In this work, we were particularly curious about the transferability of the MVD attacks. In this section, we delved into the transferability among classifiers and the transferability among encoders. We conducted ablation studies to analyse the factors that may affect the transferability.

First, as for the attacks in the black-box-encoder setting, the calibration process ensured that the MVD could decode the latent variables that had similar distributions with that output by the black-box encoder. However, we would like to investigate whether we can use fewer data to query the encoder during the calibration. In other words, we wanted to find out the minimal number of data required for estimating the mean $\mu$ and the variance $\delta$ of the latent distribution. We adopted stratified sampling with a rate of $p$ to downsample the data from each data category. Given the sampled sub-dataset, we then estimated the $\mu$ and the $\delta$ of the black-box encoder and calibrated the MVD to generate adversarial examples from the distributions. 

\begin{figure}[t!]
\center
\includegraphics[width=1\linewidth]{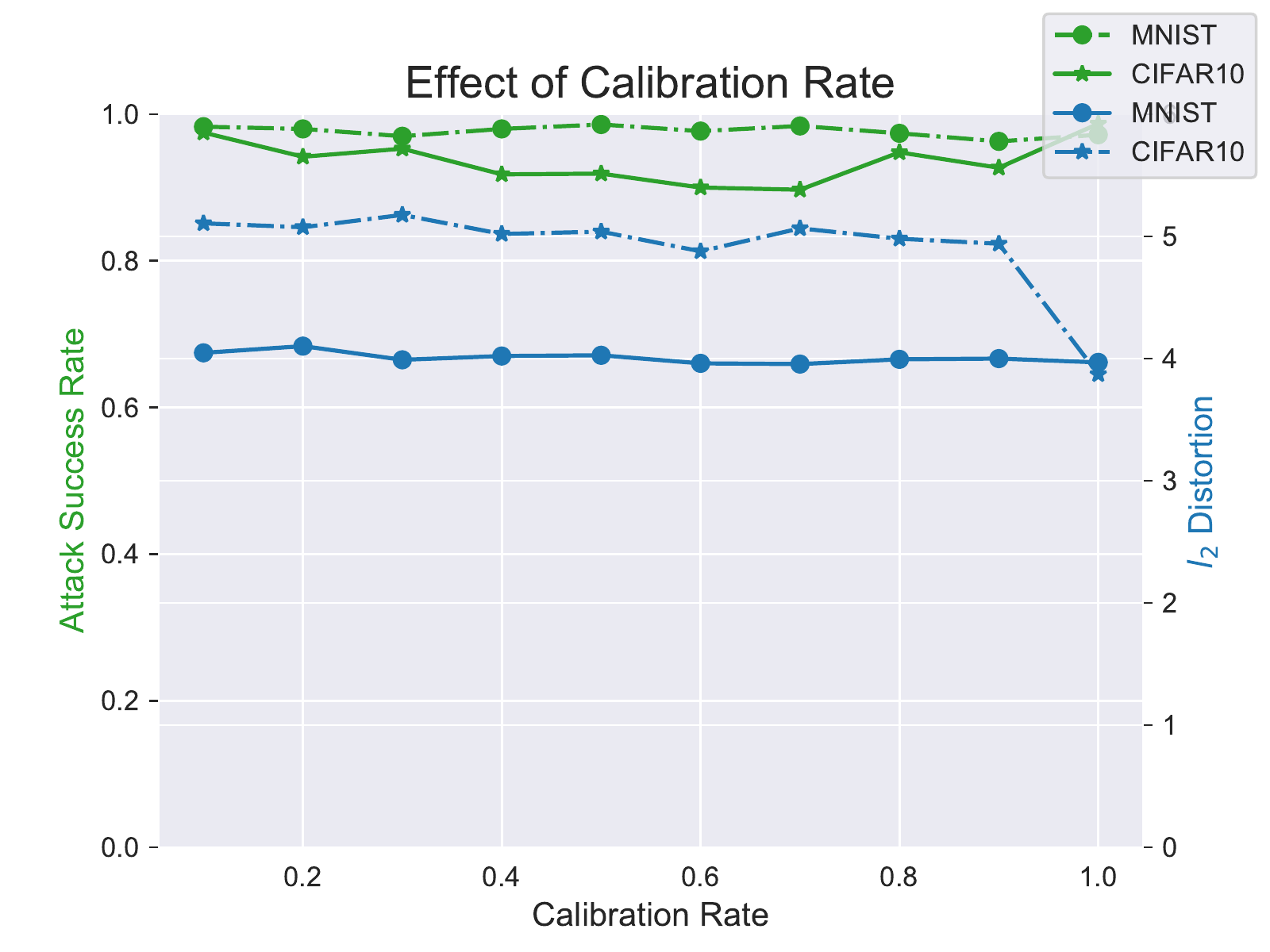}
\caption{The relationship between the attack success rate and the sampling rate used in the MVD calibration.}
\label{rate_vs_suc_dist}
\end{figure}

We empirically swept the sample rate from 0.1 to 1 every 0.1 to calibrate a set of MVDs. Next, we trained the MVDs to be adversarial against the same white-box victim classifier. Subsequently, we evaluated the attack success rate of the MVDs against the white-box victim classifier. We also evaluated the $l_2$ norm of the perturbations generated by the MVDs. The relationship among the attack success rate, the $l_2$ norm and the sampling rate was plotted in Figure \ref{rate_vs_suc_dist}. According to the plot, the sample rate (\textit{i.e.}, calibration rate) actually had no significant impact on the attack success rate and the distortion of the MNIST examples. However, we could observe that the high calibration rate could reduce the $l_2$-norm of the CIFAR10 adversarial examples. The possible reason is that, compared to that of the MNIST examples, the learnt latent distribution of the CIFAR10 examples were farther away from $\mathcal{N}(0,1)$. Henceforth, the MVD would have a high decode error if it was not well calibrated.

\begin{figure}[t!]
\center
\includegraphics[width=1\linewidth]{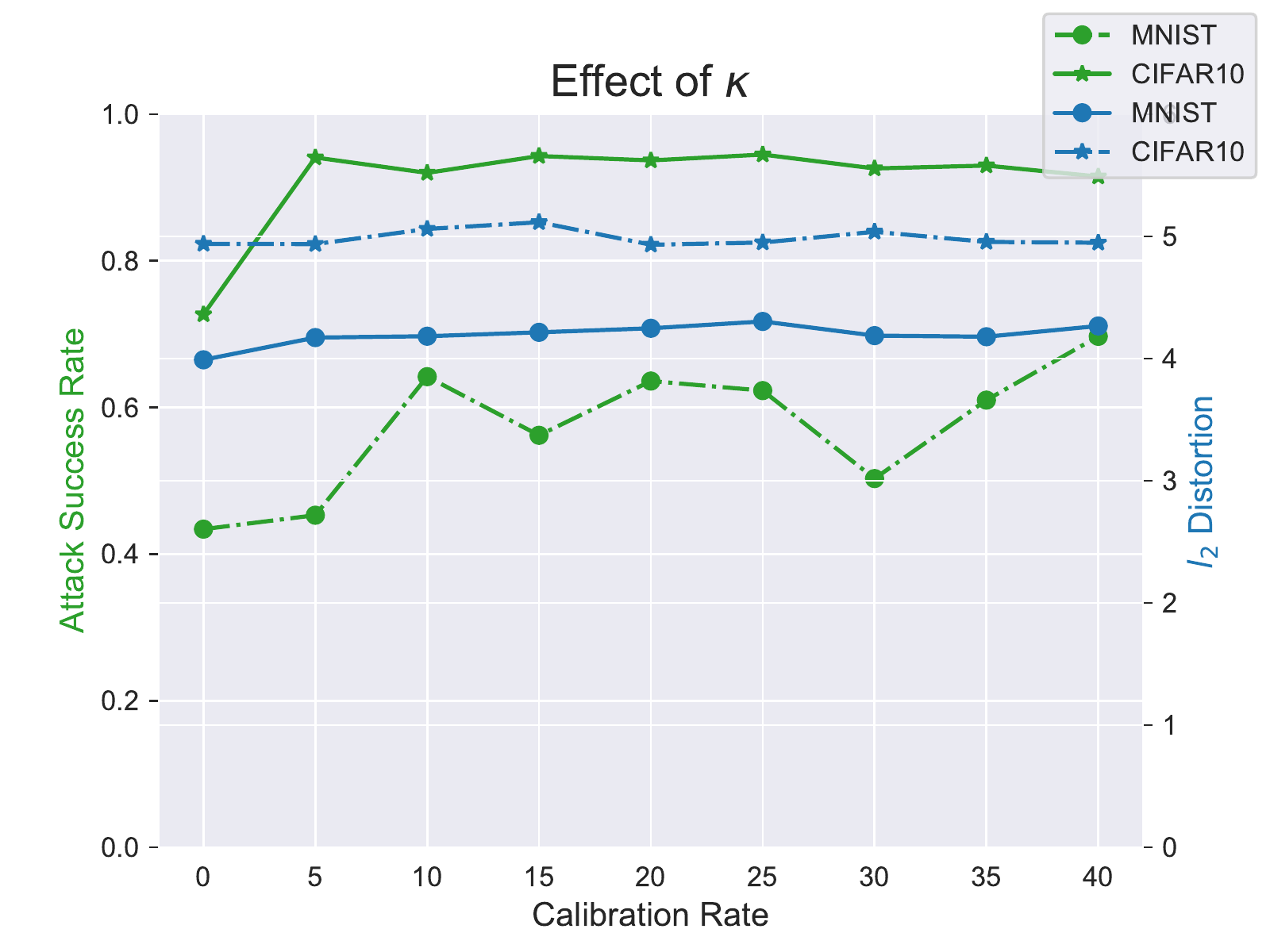}
\caption{The success rate and the $l_2$-norm of the adversarial perturbations induced by MVDs trained under different $\kappa$ values.}
\label{kappa_vs_succ_dist}
\end{figure}

The transferability of the adversarial examples among classifiers has been widely evaluated in previous works \cite{liu2016delving, papernot2017practical, tramer2017space}. However, the transferability of generative models has not been thoroughly looked into. In this section, we delved into the transferability of adversarial examples generated by the MVD. According to the experiments in Section \ref{blackbox_eval} and Section \ref{whitebox_eval}, we observed that the MVD attacks can transfer among classifiers. Compared to targeted attacks, non-targeted attacks generally had better transferability among the classifiers. This observation aligns with that observed from the query-and-optimised adversarial examples \cite{liu2016delving}. Furthermore, evaluated the impact of the hyper-parameter $\kappa$ in the loss function \ref{eq4}, which controls the misclassification confidence of the generated adversarial examples. We swept $\kappa$ from 1 to 40 every 5 to observe changes in the success rate of the transferring attacks and the $l_2$-norm of the perturbations. We plotted the trends in Figure \ref{kappa_vs_succ_dist}. Being different to the C$\&$W attack, increasing $\kappa$ value in MVD did not boost the $l_2$-norm of the perturbations. However, higher $\kappa$ induced a slightly higher success rate of black-box attacks.

\section{Discussion}\label{S::Discuss}
Based on the methods introduced in Section \ref{S::Method}, we further investigate how to instantiate such an MitM attack into real-world applications. Herein, we provide a case study about how to launch the MitM attack towards cloud-based machine learning API services using MVD. Furthermore, we discuss a possible adaptive defence against the MVD.

\subsection{Case study: MitM attacks against cloud-based ML APIs}
An application that might be exploited by an MitM attacker is cloud-based ML API service. Users of the ML services usually stream their data to online DNN models by using provided API functions. For example, Google Cloud Vision APIs transmit a client's requests through HTTPS protocol to an online server for processing (\textit{e.g.}, image labelling, face detection, and logo detection). The processed results will then be returned to the client. One primary objective of MitM attacks against such a client-server system might be impersonating the server and generating incorrect returns to the client. Directly manipulating the data or the returns from the server during transmission is difficult since 1) the data is encrypted and 2) the possible access points (\textit{e.g.}, routers) to the transmitted data are protected. Alternatively, the attacker may manipulate the data within the client machine and generate returns locally. However, this attack can be easily detected by monitoring the inbound/outbound network traffic of the client machine.

As a more stealthy means of launching MitM attacks, the attacker can plant an adversarial generative model into the client terminal. The attacker can fake the Ml APIs and redirect the client requests to the adversarial generative model. Later, the data processed by the adversarial generative model will be submitted to the online server for processing. There are several pathways for achieving this (\textit{e.g.}, trojaned API client distributions, malware, and DLL injections). Then, the attacker can first redirect the API requests to the local adversarial model. The adversarial model will process the data examples in the requests locally and then forward them to the online server. This attack vector has several advantages. First, the attacker can complete attacks without accessing other devices (\textit{e.g.}, routers) between the client and the server. Second, the attacker does not need to decrypt the transmitted data traffic. Third, unlike directly replacing data examples submitted by the client, the attacker is not required to inspect the details of the data before launching the attacks. At last, the attacks are long-persistent threats, instead of one-off attacks.

\subsection{Adaptive defence}
According to the definition of the problem in Section \ref{S::Problem_def}, the attacker acts as an insider who designs and plants malicious generative models into the victim model pipeline prior to the development of the model. Here we discuss possible adaptive defence means towards two scenarios. In the first scenario, the attacker secretly inserts a malicious VAE between input data and a classifier. In the second scenario, the attacker replaces a VAE decoder of a VAE-classifier structured model to the MVD . 

In the case of inserting a malicious VAE, we assume that model users are not aware of the existence of the VAE. A possible defence is to apply randomisation techniques to the model input, such that the adopted random transformations will mitigate the effect of the adversarial perturbations. However, this defence cannot be kept intact at all the time. As suggested by several studies about query-optimised adversarial examples, it is possible to optimise the adversarial examples over an expectation of transformations to generate robust examples that can defeat the randomisation defence \cite{Athalye2017, obfuscated-gradients}. Theoretically, adversarial generative models can also optimise themselves against a set of transformations to achieve the robustness against the defence. Proactive defences can also be applied here. As a defence, the DNN classifier itself will be trained by adversarial gradients or robust optimisation techniques (\textit{e.g.}, Lipschitz continuity), such that it is insensitive to adversarial perturbations \cite{tramer2017ensemble, madry2017towards}. These methods can secure the classifier from black-box attacks. However, many proactive defences are still vulnerable to white-box adversaries. 

If the victim model has a VAE-classifier structure, the VAE inputs might not be accessed by the classifier. For instance, VAE encoder and decoder have been used for compressing data in sensor networks \cite{liu2019cbn}. A data receiver, in this case, cannot obtain the inputs to the encoder. Henceforth, simultaneously monitoring the classification results on the input and the output from the VAE might not be an applicable defence. The proactive defence methods applied to the classifier can partially foil adversarial attacks. However, they still suffer from white-box adversaries. 

\section{Conclusion}\label{S::Conclude} 
In this paper, We propose a Malicious VAE Decoder (MVD) for generating adversarial examples of DNN classifiers. First, the MVD can be used together with VAE encoders as a generative model that produce data-agnostic adversarial examples in real-time. Second, in VAE-classifier structured machine learning systems, the MVD can decode the input data examples of benign black-box VAE encoders to their adversarial counterparts. In both cases, the evaluation results suggest that the MVD can produce adversarial examples that lead to misclassification of the DNN classifiers with high success rates. This enhances the adaptability of MitM adversaries.

In this study, we adopted a simple VAE architecture to study the adversarial example generation problem. As a drawback of the method, the fidelity of the adversarial example is affected by the VAE reconstruction quality. To address this problem, one may use architectures such as VAE-GAN to enhance the reconstruction fidelity. In our future work, we will first study the methods (\textit{e.g.}, trojaned generative models) for making the MitM attacks more stealthy. Second, we will further study the possible threats brought by adversarial generative models on real-world ML applications. Last but not least, we will implement the deliberate MitM attack against real-world ML applications, such that we can reveal undiscovered vulnerabilities of the ML applications and propose reliable defensive methods.

\bibliographystyle{abbrv}
\bibliography{../advref}
\end{document}